\documentclass[aps,twocolumn,superscriptaddress,longbibliography,nofootinbib]{revtex4-2}
\usepackage{amsmath,amssymb}
\usepackage{amsthm}
\usepackage{graphicx}
\usepackage{bm}

\usepackage{orcidlink}
\usepackage{xcolor}
\usepackage{comment}
\providecommand{\ket}[1]{|#1\rangle}

\theoremstyle{remark}

\newcommand{\rev}[1]{\textcolor{blue}{#1}}

\begin{document}

\title{ Newtonian Gravitational Curvature-Induced Entanglement Generation}

\author{Mughees Ahmad Khan\orcidlink{0009-0009-4214-7637}}
\email{mukh68937@hbku.edu.qa}
\affiliation{Qatar Center for Quantum Computing, College of Science and Engineering, Hamad Bin Khalifa University, Doha, Qatar}

\author{Asad Ali\orcidlink{0000-0001-9243-417X}}
\email{asal68826@hbku.edu.qa}
\affiliation{Qatar Center for Quantum Computing, College of Science and Engineering, Hamad Bin Khalifa University, Doha, Qatar}

\author{M. I. Hussain~\!\!\orcidlink{0000-0002-6231-7746}}
\affiliation{Qatar Center for Quantum Computing, College of Science and Engineering, Hamad Bin Khalifa University, Doha, Qatar}

\author{H. Kuniyil~\!\!\orcidlink{0000-0003-0338-1278}}
\affiliation{Qatar Center for Quantum Computing, College of Science and Engineering, Hamad Bin Khalifa University, Doha, Qatar}

\author{Saif Al-Kuwari\orcidlink{0000-0002-4402-7710}}
\email{smalkuwari@hbku.edu.qa}
\affiliation{Qatar Center for Quantum Computing, College of Science and Engineering, Hamad Bin Khalifa University, Doha, Qatar}

\date{\today}

\begin{abstract}
We show that gravitational curvature can control the generation of nonlocal
quantum correlations in a hybrid qubit-mechanical device. The tidal field of a
nearby source mass modifies the susceptibility of a shared mechanical
oscillator, thereby tuning an oscillator-mediated qubit-qubit interaction and
the resulting entangling phase. An exact treatment of the dynamics reveals
stroboscopic geometric gates whose accumulated phase is directly sensitive to
gravitational curvature. Treating the curvature as an unknown parameter, we
derive the ultimate quantum limit for its estimation and identify a
parity-based measurement that saturates this bound. Solving the full master equation with the
mechanical mode retained explicitly, we find that thermal occupation of the
mediator suppresses entanglement between the closure times but is undone at
each closure, exactly in the unitary limit for any initial mechanical temperature, so that ground-state cooling of the oscillator is not a prerequisite for the protocol. Mechanical
damping and qubit dephasing behave differently: they leak branch information
irreversibly to the environment, and it is the heating and dephasing rates,
rather than the bath occupation alone, that limit the entanglement
visibility and the number of usable interrogation loops. In contrast to
gravity-mediated entanglement proposals, entanglement is generated by the
mechanical oscillator while gravity acts solely as a classical control field. We quantify the achievable curvature parameters; all curvature dependencies are analytic, allowing for exact rescaling of the results. The scheme therefore demonstrates
gravitational control of a quantum interaction and provides a route to
curvature sensing based on a nonlocal two-qubit phase rather than on local
phase measurements.
\end{abstract}

\maketitle

\section{Introduction}
\label{sec:intro}

Mechanical oscillators have emerged as sensitive transducers connecting gravitational fields to
quantum observables~\cite{aspelmeyer2014cavity,degen2017quantum}. Their center-of-mass motion couples
directly to Newtonian forces, and advances in levitation, cooling, and quantum control have pushed
nanoscale resonators into regimes where quantum effects are measurable even under gravitational
perturbations~\cite{gonzalezballestero2021levitodynamics}. The gravitational field of millimeter-sized source masses has been resolved \cite{westphal2021measurement}, and a mechanical resonator has been prepared in its quantum ground state and controlled at the single-phonon level \cite{oconnell2010quantum}. A mechanical resonator is therefore a natural interface at which a gravitational field can potentially be converted into a quantum mechanical observable.

Most quantum gravimetry protocols encode gravitational information in a local observable; an acceleration-induced phase or a frequency shift of a single resonator \cite{kasevich1991atomic, qvarfort2018gravimetry}. The estimator is then built from a single probe, and entanglement plays no role in the encoding; the precision limits of this single-probe approach have been derived at the level of quantum Fisher information \cite{armata2017quantum}. Entanglement and nonlocal correlations can, in principle, improve sensitivity beyond
classical limits and extract information inaccessible to single-probe measurements~\cite{giovannetti2011advances,braunstein1994statistical,paris2009quantum}.
This raises the question of whether a gravitational field can be transduced into a measurable
two-body entangling interaction within a hybrid quantum device.

We study a hybrid architecture consisting of two qubits coupled longitudinally to a common mechanical resonator. Longitudinal coupling commutes with the qubit
Hamiltonian, admits exact analytical treatment, and produces high-fidelity oscillator-mediated
gates~\cite{royer2017fast,didier2015fast}. The resonator susceptibility determines the strength of the induced two-qubit Ising
interaction; gravitational curvature modifies this susceptibility and thereby controls the rate
of entanglement generation. The shared mode is essential because the displacement conditioned on the state of one qubit acts on the other qubit, and eliminating the oscillator leaves a two-qubit interaction. The individual components of geometric phase gates, facilitated by a shared bosonic mode, and the softening of a mechanical frequency due to Newtonian effects are both well-established concepts. This work presents a significant contribution by assembling a fully analytical transduction chain that connects the tidal field to the entangling phase of two qubits. Additionally, it provides a precise analysis of which parts of this chain remain intact despite thermal occupation, mechanical damping, and qubit dephasing.

The mechanism is conceptually distinct from the Bose-Marletto-Vedral (BMV) proposals~\cite{bose2017spin,marletto2017gravitationally},
in which the gravitational field itself is the candidate quantum mediator. In BMV  schemes,
observation of entanglement may constitute evidence for the quantum nature of gravity, and optomechanical implementations of this idea have been analyzed in detail \cite{matsumura2020gravity,krisnanda2020observable}; gravity-mediated entanglement between two oscillators has been analyzed as a quantum superposition of geometries \cite{bengyat2024}, and the source-mass form factors that optimize gravity-induced entanglement have been derived \cite{tang2025}. In the present work the mechanical oscillator, not gravity, is the explicit quantum mediator: gravity does not transmit quantum information between the qubits but modifies a property of the mediator, its frequency, which enters only as a classical background field. The observation of entanglement here is evidence of gravitational control of a quantum gate, not of graviton
exchange; no quantization of the gravitational field is assumed. Related work on gravity-induced
decoherence~\cite{torovs2024loss}, optomechanical gravity signatures~\cite{miki2024quantum}, and quantum correlations in curved space-time~\cite{ali2024quantum} illustrates the breadth of gravitational effects accessible in quantum-mechanical systems.

The gravitational effect of primary interest in our current contribution is not the uniform acceleration due to gravity but rather \emph{the curvature,
or tidal field, associated with the spatial gradient of the gravitational force}. The equivalence principle states that a uniform field can be eliminated by transitioning to a freely falling frame, while the tidal field cannot be removed \cite{will2014confrontation,misner1973gravitation}. Uniform acceleration shifts the oscillator equilibrium and produces local single-qubit phases suppressible by a spin-echo sequence \cite{hahn1950spin}; being local unitaries, these phases cannot generate or alter entanglement. Curvature modifies the effective restoring force and mechanical frequency, changing 
the susceptibility of the shared mode and encoding a nonlocal two-qubit phase; this relative phase between two-qubit computational states is observable. It is the second-order spatial gradient of the Newtonian potential that enters the two-qubit dynamics.

This paper quantifies this mechanism. A source of mass \(M\) at a distance
\(R\) produces a bare curvature \(\Gamma_R = GM/R^3\). The linear gravitational
force displaces the oscillator equilibrium, so the curvature that acts on the
mode is \(\Gamma_\mathrm{eq} = GM/(R + x_\mathrm{eq})^3\), which renormalizes the
mechanical frequency from its bare value \(\omega_0\) to \(\Omega^2 = \omega_0^2 - 2\Gamma_\mathrm{eq}\). The renormalized frequency sets the susceptibility of the shared mode and hence the strength of the induced Ising coupling \cite{ali2024ergotropy,ali2024trade}. We treat the qubit-oscillator coupling exactly; a polaron
transformation removes the coupling, and the resulting Magnus expansion terminates at second order, so the evolution operator holds at arbitrary coupling strength without rotating-wave or dispersive approximations~\cite{magnus1954exponential,blanes2009magnus}. At stroboscopic times, the oscillator disentangles from the qubits and leaves a pure two-qubit phase gate, in direct analogy with the geometric phase gates of trapped-ion quantum computing \cite{sorensen2000entanglement,leibfried2003experimental}. We then retain the mechanical mode explicitly in a Markovian master equation with mechanical damping, a thermal bath, and qubit dephasing. We show that the effect of an initially thermal mediator is undone exactly at the closure times, whereas damping and dephasing leak information about the qubit state to the environment and limit the usable interrogation time. Finally, treating the curvature as an unknown parameter of the closed-system state at closure, we derive the quantum Fisher information, propose a parity-measurement readout that saturates the quantum Cram\'er--Rao bound, and add a differential source protocol; the Fisher-information analysis of the open-system state is left for future work.

The paper is organized as follows. Section~\ref{sec:model} introduces the Hamiltonian and the curvature-dependent mechanical frequency. Section~\ref{sec:interaction} derives the effective qubit-qubit interaction. Section~\ref{sec:entanglement} analyzes the entanglement dynamics and the stroboscopic entangling gate. Section~\ref{sec:open} treats the open-system dynamics with the mechanical mode retained explicitly and separates the reversible effect of an initially thermal oscillator from the irreversible effects of mechanical damping and qubit dephasing. Section~\ref{sec:metrology} formulates the curvature estimation problem for the closed-system state at closure, derives the quantum Fisher information, and presents the parity-based readout together with \rev{the spin-echo-assisted differential sensing scheme}. Section~\ref{sec:summary} concludes, and the detailed derivations are provided in Appendices~\ref{app:renorm}--\ref{app:qfi}.

\section{Model and Gravitational Renormalization}
\label{sec:model}
\subsection{Hamiltonian}

\begin{figure}[t]
  \centering  \includegraphics[width=\columnwidth]{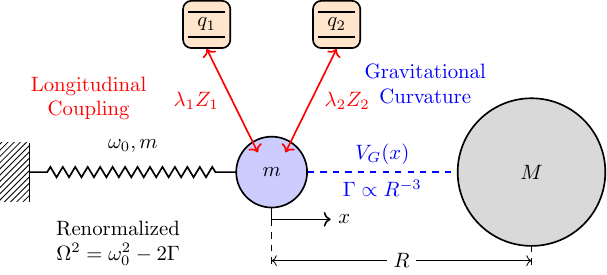}
  \caption{Schematic of the proposed setup. A mechanical oscillator of mass $m$ and bare frequency $\omega_0$ is placed at a distance $R$ from a source mass $M$. The Newtonian gravitational potential $V_G(x)$ induces a curvature $\Gamma \propto R^{-3}$ which renormalizes the mechanical frequency to $\Omega^2 = \omega_0^2 - 2\Gamma$. Two qubits ($q_1, q_2$) are coupled longitudinally to the oscillator with strengths $\lambda_1, \lambda_2$. The gravitational curvature controls the effective susceptibility of the oscillator, thereby modulating the mediated two-qubit entanglement rate. The displacement $x$ is measured along the line joining the oscillator and the source.}
  \label{fig:schematic}
\end{figure}

Our system consists of two qubits with transition frequencies $\omega_{q,i}$ ($i = 1,2$) coupled
longitudinally to a harmonic mechanical mode of bare frequency $\omega_0$ and mass $m$ as shown in Fig.~\ref{fig:schematic}.
A source mass $M$ is at distance $R$ from the mechanical equilibrium. The total Hamiltonian is
\begin{equation}
  H = \hbar\omega_0 b^\dagger b
    + \sum_{i=1}^{2}\frac{\hbar\omega_{q,i}}{2}Z_i
    + \hbar\sum_{i=1}^{2}\lambda_i Z_i(b+b^\dagger)
    + V_G(x),
  \label{eq:H_total}
\end{equation}
where $\hbar$ is the reduced Planck constant, $b$ ($b^\dagger$) is the annihilation (creation) operator of the phonon mode, $\omega_0$ and $\omega_{q,i}$ are the angular frequencies of the oscillator and of the qubits, $Z_i$ is the Pauli-$z$ operator of qubit $i$ with $Z_i\ket{0}_i=+\ket{0}_i$, $\lambda_i$ is the longitudinal coupling rate, $V_G(x) = -GMm/(R+x)$ is the Newtonian interaction with $G$ Newton's constant, and $x = x_\mathrm{zpf}(b+b^\dagger)$ is the displacement of the oscillator along the line joining it to the source, with $x_\mathrm{zpf}$ the zero-point amplitude given below Eq.~\eqref{eq:Omega_sq}.
The longitudinal coupling satisfies $[Z_i,\,H_q + H_\mathrm{int}] = 0$, where $H_q$ and $H_\mathrm{int}$ denote the second and third terms of Eq.~\eqref{eq:H_total}, so no excitation is exchanged with the oscillator; instead, each qubit state exerts a different force, of magnitude $\hbar\lambda_i/x_\mathrm{zpf}$, on the resonator. The qubit frequencies drop out of every result below because $Z_i$ commutes with the rest of $H$. Three assumptions are built into Eq.~\eqref{eq:H_total} and used throughout; the oscillator moves along a single axis pointing at the source, the source is static during each interrogation, and the qubits couple only through $Z_i$ (no transverse component); the last is what makes the model exactly solvable in Sec.~\ref{sec:interaction}. Apart from the expansion of $V_G$ to second order in $x$, made in Sec.~\ref{sec:renorm}, no further approximation enters the closed-system treatment of Secs.~\ref{sec:model}--\ref{sec:entanglement} and \ref{sec:metrology}; dissipation is added in Sec.~\ref{sec:open}.

\subsection{Frequency renormalization}
\label{sec:renorm}

Expanding $V_G(x)$ through Taylor series for $|x|\ll R$ as shown in  Appendix~\ref{app:renorm}, one can get
\begin{equation}
  V_G(x) \approx \mathrm{const} + mg_R x - m\Gamma_R x^2 + \mathcal{O}(x^3),
\end{equation}
where $g_R = GM/R^2$ is the gravitational acceleration of the source at the oscillator, $\Gamma_R = GM/R^3$ is the bare tidal curvature, and the constant, $-GMm/R$, only fixes the energy reference and is dropped below. The linear term displaces the equilibrium to $x_\mathrm{eq}\simeq -g_R/\omega_0^2$, and the quadratic term reduces the effective stiffness of the oscillator: the side nearer to the source is pulled more strongly than the far side. The neglected cubic term is smaller than the quadratic one by a factor of order $x/R$, below $10^{-6}$ for subnanometer displacements and millimeter distances. Expanding about the true equilibrium yields the gravity-renormalized frequency
\begin{equation}
  \Omega^2 = \omega_0^2 - 2\Gamma_\mathrm{eq},
  \label{eq:Omega_sq}
\end{equation}
where $\Gamma_\mathrm{eq} = GM/(R+x_\mathrm{eq})^3$ is the curvature at the displaced equilibrium; below we write $\Gamma$ for $\Gamma_\mathrm{eq}$ wherever the distinction from $\Gamma_R$ is immaterial. Equation~\eqref{eq:Omega_sq} is the stiffness at the displaced equilibrium and is exact for a pointlike or spherical source within the harmonic approximation. The dimensionless curvature parameter
$\epsilon \equiv \Gamma_\mathrm{eq}/\omega_0^2$ gives $\Omega = \omega_0\sqrt{1-2\epsilon}$;
the mode is stable for $\epsilon < 1/2$. In Fig.~\ref{fig:gravity_dynamics}, $\epsilon$ is set to $0$--$0.05$ for illustration; these values are chosen for visibility. The zero-point displacement, $x_\mathrm{zpf} = (\hbar/2m\Omega)^{1/2}$, is the amplitude of the ground-state phonon mode and grows as the mode softens. Note that due to the
traceless nature of the Newtonian tidal tensor in vacuum, transverse mechanical modes are
correspondingly stiffened by $+\Gamma_\mathrm{eq}$ (Appendix~\ref{app:tidal}), a unique tensor
signature of the gravitational field.

\subsection{Force and curvature channels}

The uniform acceleration $g_R$ displaces the equilibrium and produces local
qubit-frequency shifts proportional to $x_\mathrm{eq}$, and hence to $g_R$.
They arise because writing $x=x_\mathrm{eq}+X$ in the coupling term of
Eq.~\eqref{eq:H_total} shifts $b+b^\dagger$ by the constant
$x_\mathrm{eq}/x_\mathrm{zpf}$; being local unitaries, they are absorbed into
$\omega_{q,i}$ and canceled by the echo sequence of Sec.~\ref{sec:metrology}.
The curvature $\Gamma_\mathrm{eq}$ modifies $\Omega$ and the mechanical
susceptibility $\chi_m=1/m\Omega^2$, the static displacement per unit force.
Since the oscillator-mediated qubit-qubit interaction depends on $\chi_m$,
curvature is the channel encoding the nonlocal two-qubit phase.

After shifting to the fluctuation coordinate $X = x - x_\mathrm{eq}$, retaining the harmonic
approximation, the model becomes
\begin{equation}
  \frac{H}{\hbar} = \Omega b^\dagger b
    + \sum_{i=1}^{2}\frac{\omega_{q,i}}{2}Z_i
    + \sum_{i=1}^{2}\lambda_i Z_i(b+b^\dagger),
  \label{eq:H_shifted}
\end{equation}
where all parameters are evaluated at the gravitationally shifted equilibrium, and $b$ now denotes the annihilation operator of the renormalized mode of frequency $\Omega$, so that $x_\mathrm{zpf}$, and in the microscopic model of Sec.~\ref{sec:interaction}, the couplings $\lambda_i$, depend on $\Omega$. The first two terms constitute the free Hamiltonian $H_0$, and the last term is the qubit--oscillator coupling. Eq.~\eqref{eq:H_shifted} is thus Eq.~\eqref{eq:H_total} with the linear force absorbed into $x_\mathrm{eq}$ and the local qubit shifts, the curvature absorbed into $\Omega$, and the constant $V_G(x_\mathrm{eq})$ dropped.

\section{Effective Qubit-Qubit Interaction}
\label{sec:interaction}

\subsection{Polaron transformation}

The Hamiltonian~\eqref{eq:H_shifted} is exactly diagonalizable via a spin-dependent displacement.
We define the collective force operator $S = \lambda_1 Z_1 + \lambda_2 Z_2$ and the polaron unitary
$U_P = \exp[(b^\dagger - b)S/\Omega]$. Because $[S,b] = [S,Z_i] = 0$, the Baker-Campbell-Hausdorff
series terminates at first order, giving $U_P b\, U_P^\dagger = b - S/\Omega$. The linear coupling
then cancels exactly (Appendix~\ref{app:polaron}), yielding
\begin{equation}
  H' = \hbar\Omega b^\dagger b
    + \sum_{i=1}^{2}\frac{\hbar\omega_{q,i}}{2}Z_i
    - \hbar J\,Z_1 Z_2,
  \label{eq:H_prime}
\end{equation}
in which the oscillator is completely decoupled and
\begin{equation}
  J(\Gamma_\mathrm{eq}) = \frac{2\lambda_1(\Gamma_\mathrm{eq})\,\lambda_2(\Gamma_\mathrm{eq})}
                               {\Omega(\Gamma_\mathrm{eq})}.
  \label{eq:J_general}
\end{equation}
The transformation leaves a residual term $-\hbar S^2/\Omega$ (Appendix~\ref{app:polaron}), the energy gained as the oscillator relaxes to its force-dependent equilibrium. Expanding $S^2$ splits this into two parts. The third term in Eq.~\eqref{eq:H_prime} is the Ising coupling $-\hbar J Z_1 Z_2$, with $J = 2\lambda_1\lambda_2/\Omega$ the oscillator-mediated coupling strength between the qubits. The identity part, $-\hbar(\lambda_1^2+\lambda_2^2)/\Omega$, which is a global energy shift is dropped. No rotating-wave, dispersive, or weak-coupling approximation is required. The coupling $\lambda_i/\Omega$ may be arbitrary, and the result rests only on the assumptions of Sec.~\ref{sec:model} and on the absence of dissipation, lifted in Sec.~\ref{sec:open}.

\subsection{Curvature dependence}
\label{sec:curvature}

Two models for the coupling are relevant. They differ only in how $\lambda_i$
responds when curvature shifts the mode frequency $\Omega$ through
Eq.~\eqref{eq:Omega_sq}, that is, through the dimensionless curvature parameter
$\epsilon=\Gamma_\mathrm{eq}/\omega_0^2$.

In the first model the couplings are stabilized. The $\lambda_i$ are held at
their zero-curvature values $\lambda_i^{(0)}$ by calibration or feedback while
$\Omega$ varies. Only the explicit $\Omega^{-1}$ of Eq.~\eqref{eq:J_general}
then carries the curvature dependence. The stabilized coupling therefore obeys
$J_\mathrm{st}\propto\Omega^{-1}$, so that
$J_\mathrm{st}/J_\mathrm{st}(0)=(1-2\epsilon)^{-1/2}$, where $J_\mathrm{st}(0)$
is its value at $\epsilon=0$. To first order in $\epsilon$ the fractional
sensitivity is $\delta J_\mathrm{st}/J_\mathrm{st}(0)\simeq\epsilon$, with
$\delta J_\mathrm{st}=J_\mathrm{st}-J_\mathrm{st}(0)$ the curvature-induced
change. This is the model evaluated in Fig.~\ref{fig:gravity_dynamics}, where
$\lambda_1=\lambda_2$ are held fixed while $\Omega$ varies with $\epsilon$. Any
uncompensated drift of $\lambda_1\lambda_2$ is then indistinguishable from a
curvature signal of the same fractional size.

In the second model the coupling is microscopic. Each qubit couples to the
oscillator through the gradient of its transition energy, so that
$\lambda_i=\varkappa_i x_\mathrm{zpf}/2\hbar$. Here $\varepsilon_i(x)$ is the
transition energy of qubit $i$ as a function of the oscillator position, and
$\varkappa_i=\partial\varepsilon_i/\partial x|_{x_\mathrm{eq}}$ is its gradient
at equilibrium. The quantity held fixed is now $\varkappa_i$, not $\lambda_i$.
Since $x_\mathrm{zpf}=(\hbar/2m\Omega)^{1/2}$, we have
$\lambda_i\propto\Omega^{-1/2}$ and hence $\lambda_1\lambda_2\propto\Omega^{-1}$.
The microscopic coupling therefore scales as $J_\mathrm{mic}\propto\Omega^{-2}$,
which gives $J_\mathrm{mic}/J_\mathrm{mic}(0)=(1-2\epsilon)^{-1}$ and a
fractional sensitivity $\simeq 2\epsilon$, twice the stabilized-coupling result.
The enhancement arises because curvature modifies both the mechanical
susceptibility and the zero-point amplitude, each entering the coupling
independently. Equivalently, inserting $x_\mathrm{zpf}^2=\hbar/2m\Omega$ into
Eq.~\eqref{eq:J_general} gives $J=\varkappa_1\varkappa_2\chi_m/4\hbar$ with
$\chi_m=1/m\Omega^2$, so that $J$ is the product of the two force gradients and
the static susceptibility.

The general logarithmic sensitivity $J^{-1}dJ/d\Gamma_\mathrm{eq}$ follows from
Eqs.~\eqref{eq:Omega_sq} and \eqref{eq:J_general}. Differentiating
$\Omega^2=\omega_0^2-2\Gamma_\mathrm{eq}$ gives
$d\Omega/d\Gamma_\mathrm{eq}=-1/\Omega$. For $J_\mathrm{st}\propto\Omega^{-1}$
this yields $J^{-1}dJ/d\Gamma_\mathrm{eq}=1/\Omega^2$, and for
$J_\mathrm{mic}\propto\Omega^{-2}$ it yields $2/\Omega^2$. Both reduce to the
fractional sensitivities $\epsilon$ and $2\epsilon$ quoted above, because
$\Omega^2\to\omega_0^2$ and $\Gamma_\mathrm{eq}/\omega_0^2\to\epsilon$ in the
small-curvature limit.

The two models are the limits $\eta_i=0$ and $\eta_i=-1/2$ of a single
expression, where $\eta_i\equiv d\ln\lambda_i/d\ln\Omega$ measures how the
coupling responds to a change in the mode frequency. In general
$J^{-1}dJ/d\Gamma_\mathrm{eq}=(1-\eta_1-\eta_2)/\Omega^2$.

The curvature-transduction chain is
\begin{equation}
  \Gamma_\mathrm{eq}\to\Omega(\Gamma_\mathrm{eq})\to\lambda_i(\Gamma_\mathrm{eq})
  \to J(\Gamma_\mathrm{eq})\to\phi_{12}(\Gamma_\mathrm{eq}),
  \label{eq:chain}
\end{equation}
where $\phi_{12}$ is the phase accumulated on $Z_1Z_2$. It equals $\theta_k$ at
the closure times of Sec.~\ref{sec:entanglement}. Each arrow is an identity of
the harmonic model and carries no further approximation.

\section{Entanglement Dynamics}
\label{sec:entanglement}
\subsection{Exact laboratory-frame propagator}

In the interaction picture with respect to the free Hamiltonian, we conjugate the Schr\"odinger-picture coupling $\hbar S(b+b^\dagger)$ with the free propagator $U_0(t)=e^{-iH_0 t/\hbar}$, giving $H_I(t)=U_0^\dagger(t)\,\hbar S(b+b^\dagger)\,U_0(t)$. The operator $S$ commutes with $H_0$ and is therefore left unchanged, while the mode operators rotate as $U_0^\dagger(t)\,b\,U_0(t)=b\,e^{-i\Omega t}$ and its Hermitian conjugate, so the coupling reads $H_I(t)=\hbar S(b\,e^{-i\Omega t}+b^\dagger e^{i\Omega t})$. The propagator then terminates exactly at second Magnus order (Appendix~\ref{app:magnus}), because the commutator $[H_I(t_1),H_I(t_2)]\propto S^2$ commutes with all subsequent commutants:
\begin{equation}
  U_I(t) = D\!\left[\alpha(t)S\right]\exp\!\left[i\Phi(t)S^2\right],
  \label{eq:UI_exact}
\end{equation}
where $D[\beta]=\exp(\beta b^\dagger-\beta^* b)$ is the displacement operator, $\alpha(t) = (1-e^{i\Omega t})/\Omega$ is the conditional displacement amplitude, and $\Phi(t) = (\Omega t - \sin\Omega t)/\Omega^2$ is the geometric phase function. For a computational state $|s_1 s_2\rangle$, with $s_i=\pm1$ the eigenvalues of $Z_i$ and $S_s = s_1\lambda_1 + s_2\lambda_2$ the corresponding eigenvalue of $S$, the mechanical mode traces a circle of radius $|S_s|/\Omega$ in phase space. Eq.~\eqref{eq:UI_exact} is thus a qubit-state-dependent displacement times a qubit-state-dependent phase, $\Phi(t)S_s^2$ being proportional to the phase-space area swept by the branch $\ket{s_1s_2}$; this is the structure of the M{\o}lmer-S{\o}rensen and geometric phase gates of trapped ions~\cite{sorensen2000entanglement,leibfried2003experimental}, realized here in the limit in which the state-dependent force is static in the laboratory frame; closely related geometric phase gates based on ac-Stark shifts have been formulated for cavity-QED platforms \cite{hussain2014geometric,hussain2015geometric}. In the trapped-ion gates, the phase-space loop is driven by a bichromatic field and closes at the inverse detuning; here the coupling in Eq.~\eqref{eq:H_shifted} carries no explicit time dependence, the residual rotation in the interaction picture occurs at the mechanical frequency itself, and the loop closes at $t_k=2\pi k/\Omega$.
\begin{figure*}[t]
  \includegraphics[width=0.8\textwidth]{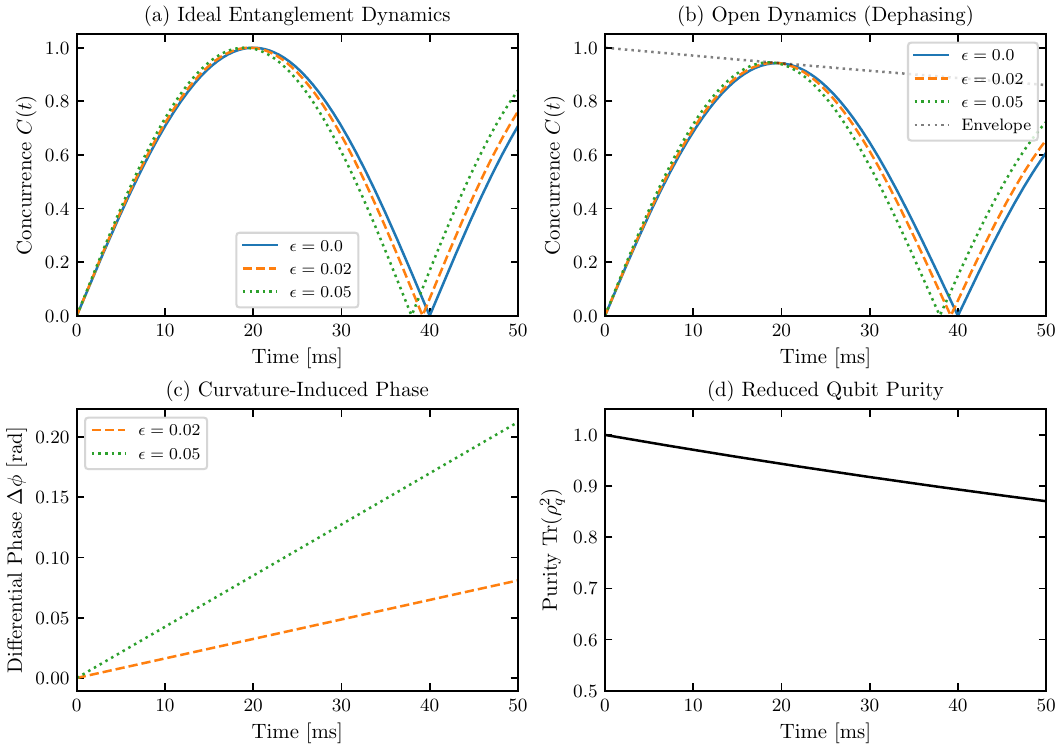}
  \caption{Effective-model entanglement dynamics as a function of gravitational curvature.
    (a)~Two-qubit concurrence for $\epsilon = 0$, $0.02$, and $0.05$ with no dephasing.
    Increasing $\epsilon$ shifts the oscillation via the enhanced susceptibility.
    (b)~Same as (a) with independent qubit dephasing $\gamma_\phi = 1.5\ \mathrm{s^{-1}}$;
    the envelope decays as $e^{-2\gamma_\phi t}$.
    (c)~Differential entangling phase $\Delta\phi = 2t[J(\epsilon)-J(0)]$ versus time,
    showing linear accumulation of the curvature-induced phase difference.
    (d)~Reduced two-qubit purity $\mathrm{Tr}(\rho_q^2)$ under qubit dephasing only. Purity
    curves for the three $\epsilon$ values overlap in this panel because the dephasing channel
    is $\epsilon$-independent in the effective model; the small spread becomes visible at late
    times due to the $\epsilon$-dependent oscillation frequency.
    Parameters arbitrarily set are $\omega_0/2\pi = 200\ \mathrm{Hz}$, $\lambda/2\pi = 25\ \mathrm{Hz}$,
    $J(0)/2\pi \simeq 6.25\ \mathrm{Hz}$.}
  \label{fig:gravity_dynamics}
\end{figure*}

\subsection{Stroboscopic closure}

The trajectories close, $\alpha(t_k) = 0$, at $t_k = 2\pi k/\Omega$ for $k = 1,2,3,\ldots$
At these times, the displacement operator becomes the identity, and the oscillator disentangles
from the qubits exactly, regardless of the initial mechanical state (Sec.~\ref{subsec:thermal} makes this explicit for a thermal initial state). The Schr\"odinger-picture
mechanical state returns to its initial condition since $e^{-i\Omega t_k b^\dagger b} = \mathbf{1}$
for integer $k$. As a result, the mechanical mode factors out of the joint state. The qubits remain in a pure two-qubit state, disentangled from the oscillator, while the qubit-qubit correlations generated during the loop are retained. The accumulated two-qubit gate is
\begin{equation}
  U_\mathrm{ent}(t_k) = \exp\!\left[i\theta_k Z_1 Z_2\right],
  \quad \theta_k = \frac{4\pi k\lambda_1\lambda_2}{\Omega^2} = Jt_k,
  \label{eq:Uent}
\end{equation}
where $\theta_k$ is the qubit-state-dependent geometric phase after $k$ complete loops projected onto $Z_1Z_2$, the global phase $\Phi(t_k)(\lambda_1^2+\lambda_2^2)$ has been dropped (Appendix~\ref{app:magnus}), and the second equality confirms consistency with the polaron-frame Hamiltonian. Closure does require the loop time to be set at the curvature-shifted frequency $\Omega$, a mistimed loop leaves a residual displacement $\alpha(t)S$ and hence residual qubit-oscillator correlation.

\subsection{Concurrence at closure}

For the initial separable state $\ket{+}_1\ket{+}_2$, with
$\ket{+}_i=(\ket{0}_i+\ket{1}_i)/\sqrt2$, the Wootters
concurrence~\cite{wootters1998entanglement} at a closure time is
\begin{equation}
  C(t_k) = |\sin(2\theta_k)| = |\sin(2Jt_k)|.
  \label{eq:concurrence}
\end{equation}
Maximum entanglement ($C=1$) occurs when $\theta_k=\pi/4+n\pi/2$, with $n$ a
non-negative integer. Both $J$ and $t_k$ carry the curvature dependence of
$\Omega(\Gamma_\mathrm{eq})$. The closure phase $\theta_k$ therefore encodes
$\Gamma_\mathrm{eq}$. It does so with higher sensitivity than a fixed-time
protocol, because the closure time lengthens as the mode softens. Here
$t_k\propto\Omega^{-1}$, so Eq.~\eqref{eq:Uent} gives
$\theta_k\propto\Omega^{-2}$ for stabilized couplings and
$\theta_k\propto\Omega^{-3}$ for microscopic couplings. The fractional
sensitivity $\theta_k^{-1}\partial_{\Gamma_\mathrm{eq}}\theta_k$ is then
$2/\Omega^2$ and $3/\Omega^2$, respectively. For a phase $Jt$ accumulated over a
fixed time $t$, it is $1/\Omega^2$ and $2/\Omega^2$.
 
Fig.~\ref{fig:gravity_dynamics} is computed from the effective Hamiltonian of
Eq.~\eqref{eq:H_prime}, after the oscillator has been eliminated. It is
evaluated at a fixed time, not at the closure times. Panel~(a) plots
$C(t)=|\sin 2J(\epsilon)t|$ for $\epsilon=0$, $0.02$, and $0.05$. Curvature
enters only through $J(\epsilon)$. Its fractional change $\simeq\epsilon$ shifts
the first maximum, at $t=\pi/4J(0)$, by $\simeq-\epsilon t$. Panel~(c) plots the
same effect as the phase difference $\Delta\phi=2t[J(\epsilon)-J(0)]$. This phase
is linear in $t$ because $J$ is time-independent. Panels~(b) and~(d) include
qubit dephasing phenomenologically. The concurrence is multiplied by the
envelope $e^{-2\gamma_\phi t}$, which is a two-qubit coherence that decays at
twice the single-qubit dephasing rate $\gamma_\phi$. The reduced purity
$\mathrm{Tr}\rho_q^2$ of the two-qubit state $\rho_q$ follows from the same
dephasing model: the four computational-basis populations are constant, the
eight coherences with a single flipped qubit decay at $\gamma_\phi$, and the
four with both qubits flipped decay at $2\gamma_\phi$, giving
$\mathrm{Tr}\rho_q^2=\tfrac14\bigl(1+e^{-2\gamma_\phi t}\bigr)^2$, which
decays from unity toward the fully dephased value $1/4$. Both are $\epsilon$-independent by construction, so panel~(d)
contains a single curve. The stroboscopic result of Eq.~\eqref{eq:Uent}
coincides with panel~(a) at the closure times $t_k=2\pi k/\Omega$. For the
parameters of the figure, $t_1=1/(200\,\mathrm{Hz})=5\,\mathrm{ms}$. The first
maximum, $C=1$ at $t=\pi/4J(0)=20\,\mathrm{ms}$, is the closure $k=4$, with
$\theta_4=\pi/4$. The $\epsilon=0$ curve is the control for the curvature
effect. The signal is the phase difference $\Delta\phi$, not the presence of
entanglement, which is generated equally at $\epsilon=0$.
 
In summary, gravitational curvature is transduced into an observable two-qubit
entangling phase through the oscillator-mediated interaction. At the
stroboscopic closure times $t_k$ the mechanical mode disentangles exactly from
the qubits. This leaves the pure state
$\ket{\psi(\theta_k)}=e^{i\theta_k Z_1Z_2}\ket{++}$, with concurrence
$C(t_k)=|\sin(2\theta_k)|$. Both $J$ and $t_k$ inherit the curvature dependence
of $\Omega(\Gamma)$, so the accumulated phase $\theta_k=Jt_k$ carries
information about the tidal field. Concurrence quantifies the amount of
entanglement generated. It does not characterize how precisely the curvature
$\Gamma$ can be inferred from the state. This motivates a quantum-metrological
analysis, in which $\Gamma$ is treated as an unknown parameter encoded in
$\theta_k(\Gamma)$. In Sec.~\ref{sec:metrology}, we quantify the ultimate sensitivity of the closed-system protocol using quantum Fisher information, after demonstrating in Sec.~\ref{sec:open} that the stroboscopic mechanism remains effective despite realistic noise. Furthermore, we identify a measurement that achieves the corresponding quantum limit.
 
\section{Open-System Dynamics}
\label{sec:open}

The preceding sections establish the exact unitary evolution of the
qubit-oscillator system. A realistic mechanical mode, however, exchanges
energy with a thermal reservoir, while the qubits experience phase noise.
These processes must be included to determine whether the stroboscopic
disentangling mechanism of Sec.~\ref{sec:entanglement} survives under
experimentally relevant conditions.

Throughout this section we retain the mechanical mode explicitly rather than
replacing it by the effective Ising interaction of Eq.~\eqref{eq:H_prime}.
This is essential away from the closure times, where the qubits and the
oscillator are generally correlated, and it permits a direct distinction
between three physically different effects: reversible dephasing caused by an
initially mixed mechanical state, irreversible information leakage caused by
mechanical damping, and direct qubit dephasing.

\subsection{Markovian master equation}
\label{subsec:master}

Under the Born--Markov and secular approximations, the density operator of the
two qubits and the mechanical mode satisfies
\begin{align}
  \dot\rho ={}& -\frac{i}{\hbar}[H,\rho]
   + \kappa_m(\bar n_\mathrm{th}+1)\mathcal{D}[b]\rho
   + \kappa_m\bar n_\mathrm{th}\mathcal{D}[b^\dagger]\rho \nonumber\\
   &+ \sum_{i=1}^{2}\frac{\gamma_{\phi,i}}{2}\mathcal{D}[Z_i]\rho ,
  \label{eq:master}
\end{align}
where $H$ is the Hamiltonian of Eq.~\eqref{eq:H_shifted} and
$\mathcal{D}[L]\rho = L\rho L^\dagger-\tfrac12 L^\dagger L\rho-\tfrac12\rho L^\dagger L$
is the Lindblad dissipator for a jump operator $L$. The mechanical dissipators are those in the undriven mode; the static force in Eq.~\eqref{eq:H_shifted} displaces the oscillator without altering its coupling to a bath with a flat spectral density around $\Omega$, as required by the Markov approximation. Here $\kappa_m$ is the mechanical energy-decay rate and
\begin{equation}
  \bar n_\mathrm{th}=\left[\exp(\hbar\Omega/k_\mathrm{B}T_m)-1\right]^{-1}
  \label{eq:nth}
\end{equation}
the equilibrium occupation of a mechanical bath at temperature $T_m$, with $k_\mathrm{B}$ the Boltzmann constant,
evaluated at the curvature-renormalized frequency of
Eq.~\eqref{eq:Omega_sq}. The rate $\gamma_{\phi,i}$ describes pure dephasing
of qubit $i$; it is the microscopic origin of the phenomenological envelope
used in Fig.~\ref{fig:gravity_dynamics}. The first mechanical dissipator
describes phonon loss into the bath and the second thermally stimulated
absorption; together they drive the oscillator toward the thermal state with
mean occupation $\bar n_\mathrm{th}$. The qubit dissipator leaves computational-basis populations unchanged and attenuates only coherences between states with different $Z_i$ eigenvalues (Appendix~\ref{app:open}). We first
isolate the effect of an initially thermal oscillator by setting
$\kappa_m=\gamma_{\phi,i}=0$; mechanical damping and qubit dephasing are then
treated separately.

\subsection{Reduced qubit dynamics for an initially thermal oscillator}
\label{subsec:thermal}

Let the initial state be
$\rho(0)=\rho_q(0)\otimes\rho_m^\mathrm{th}(\bar n_0)$ with
$\rho_q(0)=|{+}{+}\rangle\langle{+}{+}|$ and $\rho_m^\mathrm{th}(\bar n_0)$ the thermal state of the mechanical mode with mean occupation $\bar n_0$, written out in Appendix~\ref{app:open}. We stress that $\bar n_0$ is the
occupation of the \emph{initial} mechanical state and is in general distinct
from the bath occupation $\bar n_\mathrm{th}$ of Eq.~\eqref{eq:nth}.

Applying the exact propagator of Eq.~\eqref{eq:UI_exact} to this state and tracing out the oscillator (Appendix~\ref{app:open}) shows that each coherence $\rho^{(q)}_{\bm s\bm s'}=\langle\bm s|\operatorname{Tr}_m\rho|\bm s'\rangle$ of the reduced qubit state, where $\operatorname{Tr}_m$ is the trace over the mechanical mode, between the eigenstates $|\bm s\rangle=|s_1,s_2\rangle$ of $Z_1$ and $Z_2$, with $S_{\bm s}=s_1\lambda_1+s_2\lambda_2$ the corresponding eigenvalue of $S$, acquires the phase $e^{i\Phi(t)(S_{\bm s}^2-S_{\bm s'}^2)}$ and a real factor $\chi_{\bm s\bm s'}(t)=\operatorname{Tr}_m[D(\xi)\rho_m^\mathrm{th}]$ with $\xi=\alpha(t)\Delta S_{\bm s\bm s'}$ and $\Delta S_{\bm s\bm s'}\equiv S_{\bm s}-S_{\bm s'}$. This trace is the characteristic function of the thermal state, and the thermal attenuation of the qubit coherence, $\chi_{\bm s\bm s'}(t)\equiv\mathcal{C}^\mathrm{th}_{\bm s\bm s'}(t)$, is
\begin{equation}
  \mathcal{C}^\mathrm{th}_{\bm s\bm s'}(t)
  = \exp\!\left[-\frac{2\bar n_0+1}{2}\,|\alpha(t)|^2
      \left(\Delta S_{\bm s\bm s'}\right)^2\right].
  \label{eq:thermal_factor}
\end{equation}
The conditional displacement amplitude of Eq.~\eqref{eq:UI_exact} satisfies
\begin{equation}
  |\alpha(t)|^2
  = \frac{(1-e^{i\Omega t})(1-e^{-i\Omega t})}{\Omega^2}
  = \frac{4\sin^2(\Omega t/2)}{\Omega^2},
  \label{eq:alpha_sq}
\end{equation}
so that at the closure times $t_k=2\pi k/\Omega$ of Eq.~\eqref{eq:Uent} one
has $e^{i\Omega t_k}=1$, $\alpha(t_k)=0$, and therefore
\begin{equation}
  \mathcal{C}^\mathrm{th}_{\bm s\bm s'}(t_k)=1
  \qquad \text{for every }\bar n_0 .
  \label{eq:thermal_revival}
\end{equation}
Equation~\eqref{eq:thermal_revival} is exact for unitary evolution; an
initially thermal mechanical state suppresses qubit coherence between closure
times but causes no residual loss at an exact closed-loop time. Between closures the suppression is strongest for the coherences with the largest $|\Delta S_{\bm s\bm s'}|$. For $\lambda_1=\lambda_2=\lambda$, the $|00\rangle$--$|11\rangle$ coherence, with $|\Delta S_{\bm s\bm s'}|=4\lambda$, is attenuated most, whereas the $|01\rangle$--$|10\rangle$ coherence has $\Delta S_{\bm s\bm s'}=0$, because both branches exert the same force on the oscillator, and is never affected by it. The
entangling gate of Eq.~\eqref{eq:Uent}, the concurrence Eq.~\eqref{eq:concurrence}, and consequently the curvature estimation of Sec.~\ref{sec:metrology}, which uses the state at closure, are therefore independent of the initial mechanical temperature. This is the curvature-sensing
counterpart of the thermal robustness of the M{\o}lmer--S{\o}rensen
gate~\cite{sorensen2000entanglement}, and it removes ground-state cooling of
the mediator from the list of prerequisites for the protocol. It should not,
however, be confused with immunity to mechanical damping, which transfers
information irreversibly to the external bath.

\subsection{Mechanical damping and environmental information leakage}
\label{subsec:damping}

For a fixed qubit configuration $\bm s$ the oscillator is driven by the
c-number force $S_{\bm s}$. Its conditional first moment $\beta_{\bm s}(t)=\operatorname{Tr}_m[b\,\rho_{\bm s}(t)]$, where $\rho_{\bm s}(t)=\langle\bm s|\rho(t)|\bm s\rangle/p_{\bm s}$ is the mechanical state conditioned on $\bm s$ and $p_{\bm s}$ its population, which is constant because $Z_i$ commutes with $H$ and with every dissipator, follows from Eq.~\eqref{eq:master} (Appendix~\ref{app:open}). For two branches that start from a common mechanical state, the separation $\Delta\beta=\beta_{\bm s}-\beta_{\bm s'}$ has the squared magnitude
\begin{equation}
  |\Delta\beta(t)|^2
  = \frac{\left(\Delta S_{\bm s\bm s'}\right)^2}{\Omega^2+\kappa_m^2/4}
    \left[1+e^{-\kappa_m t}-2e^{-\kappa_m t/2}\cos\Omega t\right].
  \label{eq:dbeta_sq}
\end{equation}
Damping thus deforms the trajectory itself. In particular, at the undamped
closure times,
\begin{equation}
  |\Delta\beta(t_k)|^2
  = \frac{\left(\Delta S_{\bm s\bm s'}\right)^2}{\Omega^2+\kappa_m^2/4}
    \left(1-e^{-\kappa_m t_k/2}\right)^2,
  \label{eq:residual_displacement}
\end{equation}
which is nonzero for finite $\kappa_m$: the exact geometric closure
established in Sec.~\ref{sec:entanglement} is a property of the unitary
dynamics, and with damping the nominal closure is only approximate. The residual displacement is small: $(1-e^{-\kappa_m t_k/2})^2\simeq(\kappa_m t_k/2)^2$ is of second order in $\kappa_m t_k$. It enters the qubit coherence through Eq.~\eqref{eq:thermal_factor}, with $|\alpha(t)|^2(\Delta S_{\bm s\bm s'})^2$ replaced by $|\Delta\beta(t)|^2$. The first-order effect of damping at closure is the loss of information to the bath, derived next.

We next derive the decoherence caused by information emitted into the
mechanical reservoir. Over an infinitesimal interval $dt$, mechanical damping acts as a thermal attenuation channel that mixes the mechanical mode with a thermal bath mode (Appendix~\ref{app:open}). The two conditional amplitudes $\beta_{\bm s}$ and $\beta_{\bm s'}$ leave the bath mode in two states whose overlap is its thermal characteristic function, evaluated at the separation $\sqrt{\kappa_m dt}\,\Delta\beta(t)$. Independent bath increments multiply, so the visibility accumulated over $[0,t]$ is $e^{-\Lambda(t)}$ with
\begin{equation}
  \Lambda(t)
  = \frac{\kappa_m}{2}\left(2\bar n_\mathrm{th}+1\right)
    \int_0^t|\Delta\beta(\tau)|^2\,d\tau .
  \label{eq:Lambda_general}
\end{equation}
The environment therefore distinguishes the two qubit branches in proportion
to the squared separation of their conditional mechanical trajectories. Eq.~\eqref{eq:Lambda_general} is not only a leading-order estimate: for the Markovian model of Eq.~\eqref{eq:master} it is exact, because in the Heisenberg picture every bath mode is displaced by an amount proportional to $S_{\bm s}$, and the squared distance between the bath states of two branches, summed over all bath modes, is $\kappa_m\int_0^t|\Delta\beta(\tau)|^2d\tau$. Combining the two mechanisms, a qubit coherence at time $t$ has magnitude $\exp[-\tfrac12(2\bar n_0+1)|\Delta\beta(t)|^2-\Lambda(t)]$ when $\gamma_{\phi,i}=0$; the first factor is Eq.~\eqref{eq:thermal_factor} evaluated on the damped trajectory, and the second is the loss to the bath.

In the weak-damping regime $\kappa_m/\Omega\ll1$, and for times $\kappa_m t\ll1$, the undamped trajectory may be used inside the integral, since the prefactor in
Eq.~\eqref{eq:Lambda_general} is already first order in $\kappa_m$; that is, $|\Delta\beta(\tau)|^2=|\alpha(\tau)|^2(\Delta S_{\bm s\bm s'})^2[1+O(\kappa_m\tau)]$. Inserting Eq.~\eqref{eq:alpha_sq} into Eq.~\eqref{eq:Lambda_general} and integrating (Appendix~\ref{app:open}) gives $\Lambda(t)\propto t-\sin(\Omega t)/\Omega$. At $t=t_k=2\pi k/\Omega$ the sine term vanishes and
\begin{equation}
  \Lambda(t_k)
  = \frac{2\pi k\,\kappa_m(2\bar n_\mathrm{th}+1)
          \left(\Delta S_{\bm s\bm s'}\right)^2}{\Omega^3}
    + O\!\left[\left(\frac{\kappa_m}{\Omega}\right)^{\!2}\right].
  \label{eq:Lambda_closure}
\end{equation}
The omitted terms are smaller by a factor of order $\kappa_m t_k$. For longer times the exact form Eq.~\eqref{eq:Lambda_exact} applies; its growth rate for $\kappa_m t\gg1$ is half that of Eq.~\eqref{eq:Lambda_weak}, because the damped trajectories converge to branch-dependent equilibria separated by $|\Delta S_{\bm s\bm s'}|/(\Omega^2+\kappa_m^2/4)^{1/2}$. Damping changes the entangling phase much less than it changes the visibility. With damping, the commutator $[H_I(t_1),H_I(t_2)]$ of Appendix~\ref{app:magnus} acquires the factor $e^{-\kappa_m(t_1-t_2)/2}$, and the long-time rate at which the phase accumulates changes from $1/\Omega$ to $\Omega/(\Omega^2+\kappa_m^2/4)$. At closure, $\theta_k$ is therefore shifted only by a relative amount of order $(\kappa_m/\Omega)^2$, whereas the visibility is reduced at first order, by $e^{-\Lambda(t_k)}$.

The contrast between Eqs.~\eqref{eq:thermal_revival} and
\eqref{eq:Lambda_closure} is the central result of this section. Initial
thermal mixing produces reversible qubit-oscillator dephasing that vanishes
identically at unitary closure. Mechanical damping instead exports branch
information to the bath, leaving a residual factor $e^{-\Lambda(t_k)}$ that
no choice of interrogation time can remove. Since $\Lambda(t_k)\propto\kappa_m\bar n_\mathrm{th}$ at large occupation, the
figure of merit for the mediator is the heating rate $\dot n\simeq\kappa_m\bar n_\mathrm{th}$ rather than the bath occupation
alone.
\begin{figure*}[t]
  \includegraphics[width=0.8\textwidth]{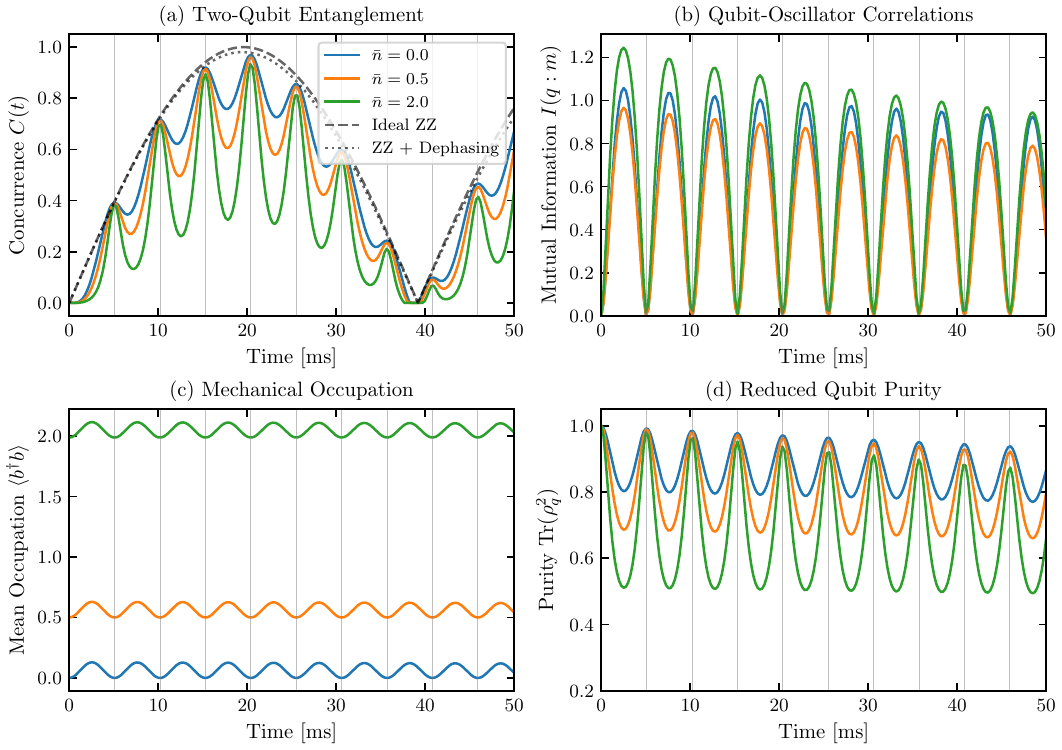}
  \caption{Open-system dynamics from the full master equation
    \eqref{eq:master}, retaining the mechanical mode explicitly.
    (a)~Two-qubit concurrence: strong suppression between closure times due to
    the thermal factor \eqref{eq:thermal_factor}, with revivals at
    $t_k=2\pi k/\Omega$ that are exact in the unitary limit and reduced by
    $e^{-\Lambda(t_k)}$ at finite $\kappa_m$ and by the dephasing factor of Eq.~\eqref{eq:dephasing_factor}.
    (b)~Qubit--oscillator mutual information $I(q{:}m)$, which vanishes at
    every closure for $\kappa_m=0$.
    (c)~Mean mechanical occupation $\bar n_m(t)$, oscillating about the
    thermal background under the spin-dependent force.
    (d)~Reduced two-qubit purity $\mathcal{P}_q(t)$, showing partial revivals
    superposed on a secular envelope set by mechanical leakage and qubit
    dephasing.
    Parameters as in Sec.~\ref{subsec:numerics}:
    $\omega_0/2\pi=200~\mathrm{Hz}$, $\lambda_{1,2}/2\pi=25~\mathrm{Hz}$,
    $\epsilon=0.02$, $\bar n_0=\bar n_\mathrm{th}=2$,
    $\kappa_m/2\pi=0.5~\mathrm{Hz}$,
    $\gamma_{\phi,1}=\gamma_{\phi,2}=0.5~\mathrm{s^{-1}}$, {and $N_b=18$ Fock states}.}
  \label{fig:open_system}
\end{figure*}

\subsection{Independent qubit dephasing}
\label{subsec:dephasing}

For a computational-basis matrix element $\rho_{\bm s\bm s'}$, the action of
the $i$th dephasing channel follows from $Z_i|\bm s\rangle=s_i|\bm s\rangle$ (Appendix~\ref{app:open}).
If $s_i=s_i'$ the matrix element is unaffected; if $s_i=-s_i'$ then
$s_is_i'=-1$ and the element decays at rate $\gamma_{\phi,i}$. A coherence
differing in the set $\mathcal{F}_{\bm s\bm s'}$ of flipped qubits therefore
acquires the factor
\begin{equation}
  \mathcal{C}^{(\phi)}_{\bm s\bm s'}(t)
  = \exp\!\left[-t\sum_{i\in\mathcal{F}_{\bm s\bm s'}}\gamma_{\phi,i}\right],
  \label{eq:dephasing_factor}
\end{equation}
which for equal rates and a coherence differing in both qubits reduces to
$e^{-2\gamma_\phi t}$. This reproduces from first principles the envelope
imposed phenomenologically in Fig.~\ref{fig:gravity_dynamics}, and unlike
Eq.~\eqref{eq:thermal_factor} it is not periodic: qubit dephasing sets the
secular decay of the signal and hence, together with
Eq.~\eqref{eq:Lambda_closure}, the largest usable loop index $k$. For the observable $Y_1Z_2$ used for readout in Sec.~\ref{sec:parity}, with $Y_i$ the Pauli-$y$ operator of qubit $i$, which connects $|\bm s\rangle$ to the state with $s_1$ reversed, $\mathcal{F}_{\bm s\bm s'}=\{1\}$ and the factor is $e^{-\gamma_{\phi,1}t}$, whereas the coherences in which both qubits are reversed carry $e^{-(\gamma_{\phi,1}+\gamma_{\phi,2})t}$, the envelope of Fig.~\ref{fig:gravity_dynamics}(b).

\subsection{Numerical implementation and convergence}
\label{subsec:numerics}

Equation~\eqref{eq:master} is solved in the Hilbert space
$\mathcal{H}=\mathcal{H}_{q_1}\otimes\mathcal{H}_{q_2}\otimes\mathcal{H}_m$ of qubit 1, qubit 2, and the mechanical mode,
with the mechanical Hilbert space truncated to the number states
$|0\rangle,\dots,|N_b-1\rangle$. Under column-stacking vectorization,
$\operatorname{vec}(A\rho B)=(B^{\mathsf T}\otimes A)\operatorname{vec}(\rho)$ for matrices $A$ and $B$,
the master equation becomes $\frac{d}{dt}|\rho\rangle\rangle=\mathcal{L}|\rho\rangle\rangle$, where $|\rho\rangle\rangle=\operatorname{vec}(\rho)$ and $\mathcal{L}$ is the Liouvillian matrix that represents the right-hand side of Eq.~\eqref{eq:master}, with formal
solution $|\rho(t)\rangle\rangle=e^{t\mathcal{L}}|\rho(0)\rangle\rangle$; the state is propagated with a sparse exponential-action algorithm. For a thermal
state the probability excluded by the cutoff is
\begin{equation}
  \epsilon_\mathrm{tail}
  = \sum_{n=N_b}^{\infty}p_n
  = \left(\frac{\bar n_0}{1+\bar n_0}\right)^{\!N_b},
  \label{eq:tail}
\end{equation}
where $p_n$ is the thermal occupation probability of the number state $|n\rangle$, Eq.~\eqref{eq:thermal_state}, which for $\bar n_0=2$ and $N_b=18$ gives
$\epsilon_\mathrm{tail}=(2/3)^{18}\simeq6.77\times10^{-4}$; convergence is verified by repeating the calculation with $N_b=25$. The illustrative
parameters are those of Fig.~\ref{fig:gravity_dynamics},
$\omega_0/2\pi=200~\mathrm{Hz}$ and $\lambda_1/2\pi=\lambda_2/2\pi=25~\mathrm{Hz}$, together with $\epsilon=0.02$, $\bar n_0=\bar n_\mathrm{th}\in\{0,\,0.5,\,2\}$, $\kappa_m/2\pi=0.5~\mathrm{Hz}$ and $\gamma_{\phi,1}=\gamma_{\phi,2}=0.5~\mathrm{s^{-1}}$. These values are chosen to display the
stroboscopic mechanism within an accessible simulation
interval; they are an idealized, platform-agnostic parameter set, and no
assignment to a specific experimental platform is implied. In every run
the bath occupation is set equal to the initial occupation,
$\bar n_\mathrm{th}=\bar n_0$, so the simulations probe the combined effect of
the two. Their distinct roles, reversible for \(\bar{n}_0\) and irreversible for \(\kappa_m \bar{n}_{\text{th}}\) are established analytically in
Secs.~\ref{subsec:thermal} and~\ref{subsec:damping}.

The reduced states are $\rho_q=\operatorname{Tr}_m\rho$ an $\rho_m=\operatorname{Tr}_q\rho$. We monitor the qubit-oscillator mutual
information $I(q{:}m)=\mathcal{S}[\rho_q]+\mathcal{S}[\rho_m]-\mathcal{S}[\rho]$ with
$\mathcal{S}[\sigma]=-\operatorname{Tr}(\sigma\ln\sigma)$ the von Neumann entropy (distinct from the force operator $S$), the
reduced purity $\mathcal{P}_q=\operatorname{Tr}[\rho_q^2]$, and the mean mechanical occupation $\bar n_m(t)=\operatorname{Tr}[b^\dagger b\,\rho_m(t)]$. The concurrence is
evaluated from $\tilde\rho_q=(Y_1Y_2)\rho_q^*(Y_1Y_2)$, with complex conjugation taken in the computational basis: if $r_1\ge r_2\ge r_3\ge r_4$
are the square roots of the eigenvalues of $\rho_q\tilde\rho_q$, then $C[\rho_q]=\max(0,r_1-r_2-r_3-r_4)$~\cite{wootters1998entanglement}, which at
closure and in the unitary limit reduces to Eq.~\eqref{eq:concurrence}.

\subsection{Open-system entanglement dynamics}
\label{subsec:open_results}

Fig.~\ref{fig:open_system} presents the full solution of Eq.~\eqref{eq:master}. Panel~(a) shows that the concurrence is strongly suppressed between closure times, because the qubits become correlated with thermally occupied, spin-dependent mechanical trajectories in accordance with Eq.~\eqref{eq:thermal_factor}. Near each nominal closure the concurrence revives and approaches the effective $ZZ$ prediction of Eq.~\eqref{eq:concurrence}. The revival is not mathematically exact at finite $\kappa_m$, owing both to the residual displacement of Eq.~\eqref{eq:residual_displacement} and to the bath-induced factor $e^{-\Lambda(t_k)}$ of Eq.~\eqref{eq:Lambda_closure}. Quantitatively, at the fourth closure ($t_4\simeq20.4$~ms) the simulated
concurrence reaches $C=0.969$ for $\bar n_0=\bar n_\mathrm{th}=0$ and
$C=0.934$ for $\bar n_0=\bar n_\mathrm{th}=2$, against $0.978$ for the
effective $ZZ$-plus-dephasing model, i.e.\ relative revival deficits of
$0.9\%$ and $4.5\%$; for the maximal coherence, Eq.~\eqref{eq:Lambda_closure}
gives $\Lambda(t_1)\simeq0.021$ at $\bar n_\mathrm{th}=2$, a visibility loss
of about $2\%$ per loop, consistent with these deficits. Closure-time values
are read at the simulation grid point nearest to $t_k$.

Panel~(b) shows the qubit-oscillator mutual information. In the unitary limit, it vanishes exactly at every closure, as required by
Eq.~\eqref{eq:Uent}. With mechanical damping its minima remain close to zero in the weak-damping regime, although exact factorization is generally lost at
finite $\kappa_m$; the correlations grow between closures as the conditional trajectories separate. Panel~(c) displays the mean mechanical occupation,
which oscillates about the thermal background because the longitudinal coupling of Eq.~\eqref{eq:H_shifted} produces spin-dependent displacements;
the small driven occupation relative to the chosen cutoff, together with the $N_b=25$ convergence test, confirms that the Fock-space truncation is
adequate. Panel~(d) shows the reduced two-qubit purity, whose partial revival near the closure times reflects the return of information temporarily stored
in the mechanical mode, and whose secular envelope is governed by irreversible mechanical leakage and direct qubit dephasing.

The simulations therefore confirm the central analytic distinction of this section: initial thermal occupation of the mediator is reversible and is undone exactly at the stroboscopic closure times, whereas mechanical damping and qubit dephasing are not. The curvature estimate of Sec.~\ref{sec:metrology} is consequently independent of the initial mechanical occupation $\bar n_0$ at closure, and is limited instead by the heating rate $\kappa_m\bar n_\mathrm{th}$ of the bath and by $\gamma_\phi$, both of which enter only through the accumulated interrogation time $t_k$. These factors multiply the coherences that carry the parity readout of Sec.~\ref{sec:parity}; their consequences for the curvature estimate are stated there.

\section{Gravitational Curvature Estimation}
\label{sec:metrology}
Throughout this section the system is closed and the state is evaluated
at exact closure, so that the oscillator is disentangled and the curvature
enters only through the phase $\theta_k(\Gamma_\mathrm{eq})$; the extension of
the estimation analysis to the open-system state of Sec.~\ref{sec:open} is
left for future work.
\subsection{Quantum Fisher information}

The curvature $\Gamma_\mathrm{eq}$ is encoded via the phase $\theta_k(\Gamma_\mathrm{eq})$.
For the input state $|{+}{+}\rangle$, the quantum Fisher information (QFI) is
(Appendix~\ref{app:qfi})
\begin{equation}
  F_Q = 4\left(\partial_{\Gamma_\mathrm{eq}}\theta_k\right)^2,
  \label{eq:QFI}
\end{equation}
and the quantum Cram\'er-Rao bound on $\nu$ independent shots is $\mathrm{Var}(\hat{\Gamma})\geq 1/(\nu F_Q)$. For stabilized couplings, $F_Q$ grows as $(\omega_0^2-2\Gamma)^{-4}$ as the mode softens; for microscopic displacement coupling as $(\omega_0^2-2\Gamma)^{-5}$ as the modes softens. Eq. (\ref{eq:QFI}) is the QFI of a single phase generated by $Z_1Z_2$, whose invariance in $\ket{++}$ is unity (Appendix {\ref{app:qfi}}; it is independent of $\theta_k$, so the amount of entanglement present at a given closure time does not determine sensitivity. The role of the two-qubit encoding is selectivity, only the nonlocal $Z_1Z_2$ phase responds to $\Gamma_{eq}$, whereas the local phases respond to $g_R$ and are removed by the echo sequence of Sec. \ref{sec:differential}.

The quantum Fisher information quantifies the maximum information about the curvature parameter $\Gamma$ that is encoded in the two-qubit state and establishes the ultimate precision permitted by quantum mechanics. However, attaining this limit requires an experimentally accessible measurement whose statistical sensitivity matches the QFI. It is therefore necessary to identify an observable that converts the curvature-dependent phase $\theta_k(\Gamma)$ into a measurable signal while preserving the full metrological advantage of the state. In the following subsection, we show that a two-qubit parity measurement provides precisely such a readout, yielding a classical Fisher information equal to the quantum Fisher information and thereby saturating the quantum Cram\'er-Rao bound.

\subsection{Parity measurement}
\label{sec:parity}

The observable $Y_1 Z_2$, with $Y_1$ the Pauli-y operator of qubit 1, satisfies (Appendix~\ref{app:qfi})
$\langle Y_1 Z_2\rangle_\Gamma = -\sin(2\theta_k)$, where $\langle . \rangle_\Gamma$ is the expectation value in the output state $\ket{\psi_\Gamma}=e^{i\theta_kZ_1Z_2}\ket{++}$, and its classical Fisher information equals $F_Q$ exactly. The two outcomes have probabilities $p_{\pm}=[1\mp sin(2\theta_k)]/2$, and the measurement saturates the quantum Cram\'er-Rao bound at every operating point for the ideal closed-system state and an ideal projective readout. Experimentally, a local $\pi/2$ pulse about $x$ on qubit 1 before standard two-qubit parity readout implements $Y_1 Z_2$.
The analytical results hold for the closed system. Qubit dephasing is included phenomenologically in Fig. (\ref{fig:gravity_dynamics}) through the envelope $e^{-2\gamma_\phi t}$, which reduces visibility of the parity fringe and therefore the information obtained per shot. Since $\theta_k$ grows with the number of loops $k$ while the envelope decays with $t_k$, the number of loops over which the entangling phase can usefully be accumulated is limited by $\gamma_\phi$, the pure dephasing rate. Mechanical dissipation is likewise not included; closure at $t_k$ is independent of the initial mechanical state, so the thermal occupation of the shared mode does not affect the stroboscopic gate, but damping of the oscillator during a loop prevents trajectories from closing exactly and leaves residual qubit-oscillator correlations at $t_k$.

\subsection{Differential Source Protocol}
\label{sec:differential}
The parity measurement described above provides an optimal readout of the
curvature-dependent phase $\theta_k$, but practical sensing also requires the
suppression of unwanted backgrounds that do not originate from the tidal field.
In particular, the uniform gravitational acceleration produced by the source
mass generates local single-qubit phase shifts through the force channel. These
phases do not carry information about the curvature and can mask the desired
entangling signal if left uncompensated.

To isolate the curvature contribution, we employ simultaneous $\pi$ pulses on both qubits, applied at stroboscopic closure times, which reverse the sign of the local $Z_i$ evolution while preserving the commuting two-qubit interaction $Z_1Z_2$ and the closed phase-space trajectories. Since the effective entangling Hamiltonian is
proportional to $Z_1Z_2$, echo pulses cancel single-body force-induced phases
without removing the accumulated entangling phase $\theta_k$. Both requirements are necessary; a $\pi$ pulse on one qubit alone maps $Z_1Z_2\to-Z_1Z_2$ and would cancel the entangling phase together with the local one, and a pulse applied away from a closure time flips the sign of $S$ in Eq.~\eqref{eq:UI_exact}, so that the trajectory no longer closes. Applied at $t_k$, the pulse pair leaves $S^2$, and hence $\theta_k$, unchanged while reversing all local terms, so that static local phases cancel over a symmetric sequence. The protocol, therefore, selectively suppresses the force channel while maintaining
sensitivity to curvature-induced modifications of the oscillator-mediated interaction.

Additional discrimination is obtained through differential source mass
modulation. By cycling the source mass between two positions, $R_A$ and $R_B$,
the experiment measures a curvature difference

\begin{equation}
\Delta\Gamma
=
GM\left(R_A^{-3}-R_B^{-3}\right),
\end{equation}
while the dominant force background varies as $R^{-2}$. Because the curvature
and force channels possess different distance scalings, modulating the source
position provides a form of spectral and geometric separation between the
desired tidal signal and residual acceleration-induced backgrounds.

The resulting differential entangling phase is

\begin{equation}
\Delta\phi_k
=
2\left(\partial_\Gamma \theta_k\right)
\Delta\Gamma ,
\end{equation}
which directly converts a change in gravitational curvature into a measurable
shift of the parity fringe, whose phase is $2\theta_k$; $\Delta\phi_k$ is the stroboscopic counterpart of the fixed-time phase difference $\Delta\phi$ of Fig.~\ref{fig:gravity_dynamics}(c). The characteristic $R^{-3}$ dependence of $\Delta\phi_k$ serves as a distinctive signature of a point-source tidal field, providing an experimental consistency check that distinguishes the curvature signal from force-channel effects and other slowly varying backgrounds. In this
way, the combination of parity readout, spin echo, and differential source modulation forms a complete sensing protocol that approaches the quantum limit
identified by the quantum Fisher information analysis.

Three points complete the protocol. First, $\Omega$ differs between the two source positions, so the closure time $t_k=2\pi k/\Omega$ must be set separately at $R_A$ and $R_B$. Second, only the differential measurement is practicable. An absolute estimate of $\Gamma_\mathrm{eq}$ from $\theta_k$ would require $\lambda_1\lambda_2$ and $\omega_0$ to be known to better than the
fractional curvature shift itself, $2\epsilon$ or $3\epsilon$
(Sec.~\ref{sec:entanglement}); source modulation instead turns slow drifts of
$\lambda_i$ and $\omega_0$ into a common-mode background. Third, the results
hold within the model of Sec.~\ref{sec:model}: a harmonic, one-dimensional
oscillator, a static pointlike or spherical source, and purely longitudinal
coupling. Damping and dephasing are treated in Sec.~\ref{sec:open}, where they
are shown to limit the largest usable loop index $k$, while the thermal
occupation of the mediator cancels at closure. The parameters of
Figs.~\ref{fig:gravity_dynamics} and~\ref{fig:open_system} are illustrative and
are not a detectability forecast. This work therefore establishes the exact
transduction chain of Eq.~\eqref{eq:chain} and its metrological
characterization under Markovian noise, not feasibility on a specific platform.

\section{Conclusion}
\label{sec:summary}

We have presented a quantum-metrological framework in which gravitational
curvature is converted into a measurable two-qubit entangling phase through
its influence on a shared mechanical oscillator. The central mechanism is the
curvature-induced renormalization of the mechanical frequency,
Eq.~\eqref{eq:Omega_sq}, which modifies the susceptibility
$\chi_m=1/m\Omega^2$ of the oscillator and thereby controls the strength of
the mediated Ising interaction between two longitudinally coupled qubits.
This establishes the explicit transduction chain of Eq.~\eqref{eq:chain},
running from the gravitational tidal field to a nonlocal quantum correlation.
 
The qubit--oscillator dynamics were treated exactly. An exact polaron
transformation eliminates the qubit--oscillator interaction without requiring
rotating-wave, dispersive, or weak-coupling approximations, yielding the
effective $ZZ$ interaction of Eq.~\eqref{eq:H_prime} whose strength
Eq.~\eqref{eq:J_general} depends directly on the curvature-renormalized
mechanical frequency. In the laboratory frame, an exactly terminated Magnus
expansion gives the propagator Eq.~\eqref{eq:UI_exact} and reveals
spin-dependent closed trajectories in mechanical phase space. At the
stroboscopic closure times $t_k=2\pi k/\Omega$ the oscillator disentangles
completely from the qubits, leaving the pure entangling gate
Eq.~\eqref{eq:Uent} with $\theta_k=Jt_k$.
 
Starting from an initially separable state, the accumulated phase generates
entanglement quantified by the concurrence of
Eq.~\eqref{eq:concurrence}, $C(t_k)=|\sin(2\theta_k)|=|\sin(2Jt_k)|$,
demonstrating that the gravitational curvature is directly encoded in an
experimentally observable two-qubit correlation. Because both the interaction
strength $J$ and the closure time $t_k$ inherit the curvature dependence of
the mechanical frequency, the resulting entangling phase provides a natural
parameter through which the tidal field can be estimated, and the closure
condition itself contributes to the sensitivity: $\theta_k\propto\Omega^{-2}$
for stabilized couplings and $\Omega^{-3}$ for microscopic displacement
coupling, against $\Omega^{-1}$ and $\Omega^{-2}$ for a phase accumulated
over a fixed time.
 
Treating the curvature as an unknown parameter, we derived the quantum Fisher
information Eq.~\eqref{eq:QFI} associated with the encoded phase and
established the corresponding quantum Cram\'er--Rao bound. We further
identified a two-qubit parity measurement whose classical Fisher information
equals the quantum Fisher information, thereby providing an optimal and
experimentally accessible readout strategy. The introduction of a
spin-echo-assisted differential source protocol enables suppression of
force-channel backgrounds while preserving sensitivity to the
curvature-dependent entangling phase, allowing the distinctive $R^{-3}$ tidal
scaling of the signal to be isolated from the $R^{-2}$ acceleration
background. These results hold for the closed system at closure; the
Fisher-information analysis of the open-system state is left for future
work.
 
The open-system analysis of Sec.~\ref{sec:open} sharpens what is and is not
required of the mediator. Solving the full master equation
Eq.~\eqref{eq:master} with the mechanical mode retained explicitly, we showed
that an initially thermal oscillator attenuates the qubit coherences by the
factor Eq.~\eqref{eq:thermal_factor}, which is periodic in the conditional
displacement and returns to unity exactly at every closure time,
Eq.~\eqref{eq:thermal_revival}, for arbitrary initial occupation
$\bar n_0$. Thermal population of the mediator is therefore reversible, and
ground-state cooling is not a prerequisite for the protocol; the entangling
gate, the concurrence, and the quantum Fisher information are all independent
of the initial mechanical temperature at closure. Mechanical damping is
qualitatively different: it exports branch information irreversibly to the
reservoir, deforming the conditional trajectories so that the nominal closure
leaves the residual displacement Eq.~\eqref{eq:residual_displacement} and
imprinting the visibility factor $e^{-\Lambda(t_k)}$ of
Eq.~\eqref{eq:Lambda_closure}, which no choice of interrogation time can
remove. Together with the qubit dephasing envelope
Eq.~\eqref{eq:dephasing_factor}, this identifies the heating rate
$\dot n\simeq\kappa_m\bar n_\mathrm{th}$ and the dephasing rate
$\gamma_\phi$, rather than the bath occupation alone, as the figures of merit
that limit the usable loop index $k$ and hence the attainable precision.
 
The scheme proposed here is conceptually distinct from gravity-mediated
entanglement proposals. The mechanical oscillator serves as the explicit
quantum mediator, while gravity acts solely as a classical perturbation that
modifies the properties of that mediator through the tidal field.
Consequently, the observation of entanglement in this setting constitutes
evidence of gravitational control of a quantum interaction rather than
entanglement generated by a quantized gravitational field.
 
Several extensions follow naturally. The trace-free structure of the
Newtonian tidal tensor (Appendix~\ref{app:tidal}) stiffens the transverse
modes by $+\Gamma_\mathrm{eq}$ while softening the radial mode by
$-2\Gamma_\mathrm{eq}$, so a two-oscillator architecture reading out both
mode families differentially would exploit a tensor signature unavailable to
scalar force sensors. Because closure must be timed at the curvature-shifted
frequency $\Omega(\Gamma_\mathrm{eq})$ rather than at $\omega_0$, adaptive
protocols that track the closure condition itself constitute a second
estimation channel worth quantifying. Finally, the analysis assumed a static
source and the harmonic truncation of Eq.~\eqref{eq:Omega_sq}; extending it
to a modulated source mass, and to the anharmonic corrections that become
relevant when the mechanical amplitude is no longer small compared with $R$,
would connect the present framework more closely to the operating regimes of
levitated and soft-clamped resonators. More broadly, the framework
demonstrates how weak gravitational curvature can be converted into a
nonlocal quantum resource and exploited for precision sensing using hybrid
quantum systems.

\section*{Data availability}
No data were created or analyzed in this study. All results are analytic
and are reproducible from the expressions in the paper together with the numerical verification.

\appendix

\section{Gravitational Renormalization of the Mechanical Mode}
\label{app:renorm}

\subsection{Newtonian expansion}

For $|x|\ll R$ the exact potential $V_G(x) = -GMm/(R+x)$ expands as
\begin{equation}
  V_G(x) = -\frac{GMm}{R}\sum_{n=0}^{\infty}(-1)^n\!\left(\frac{x}{R}\right)^n,
\end{equation}
convergent for $|x|<R$. Retaining through $x^2$ and adding the bare harmonic
potential gives $V_\mathrm{tot}^{(2)}(x) = V_0 + mg_R x
+ \frac{m}{2}(\omega_0^2-2\Gamma_R)x^2$, with $V_0$ constant,
$g_R = GM/R^2$, and $\Gamma_R = GM/R^3$.

\subsection{Equilibrium position}

Setting $dV_\mathrm{tot}/dx = 0$ for the full potential
$V_\mathrm{tot}(x)=\tfrac{m}{2}\omega_0^2x^2+V_G(x)$ gives
$u + \zeta/(1+u)^2 = 0$, with $u = x_\mathrm{eq}/R$ and
$\zeta = \Gamma_R/\omega_0^2$. Iterating $u=-\zeta(1+u)^{-2}$ for
$|\zeta|\ll 1$,
\begin{equation}
  x_\mathrm{eq} = -\frac{g_R}{\omega_0^2}\!\left[1 + 2\frac{\Gamma_R}{\omega_0^2}
    + 7\!\left(\frac{\Gamma_R}{\omega_0^2}\right)^2 + \mathcal{O}(\zeta^3)\right],
\end{equation}
so $x_\mathrm{eq}\simeq -g_R/\omega_0^2$ at leading order.

\section{Tidal Tensor in Three Dimensions}
\label{app:tidal}

The Newtonian tidal tensor $\Gamma_{ij} = -\partial_i\partial_j\Phi_G$, with
$\Phi_G = -GM/r$ the potential at distance $r$ from the source, has components
$\Gamma_{rr} = 2GM/R^3$ and $\Gamma_{\perp\perp} = -GM/R^3$ at distance $R$
from a point source, and is trace-free,
$\Gamma_{rr} + 2\Gamma_{\perp\perp} = 0$. The linearized equation of motion
along direction $i$ is $\ddot x_i=-\omega_0^2x_i+\Gamma_{ii}x_i$, so a
positive (negative) diagonal component softens (stiffens) the mode. Radially,
$\Gamma_\text{eq} = \Gamma_{rr}/2 = GM/R_\text{eq}^3$ with
$R_\mathrm{eq}=R+x_\mathrm{eq}$, giving
$\Omega_r^2 = \omega_0^2 - 2\Gamma_\text{eq}$; transversely,
$\Omega_\perp^2 = \omega_0^2 + \Gamma_\text{eq}$. The spectrum therefore
splits into one soft and two stiff modes, and the main text treats the soft
radial mode. The transverse shift is $+\Gamma_\text{eq}$ against
$-2\Gamma_\text{eq}$ for the radial mode, so a differential
radial--transverse readout carries $3\Gamma_\text{eq}$, a factor $3/2$ over
the radial mode alone~\cite{degen2017quantum}.

\section{Exact Polaron Transformation}
\label{app:polaron}

Define $A = S/\Omega$ and $K = (b^\dagger-b)A$, so that $U_P=e^{K}$ is the
displacement operator $D[A]$ of Eq.~\eqref{eq:UI_exact} with operator-valued
amplitude $A$, which commutes with $b$, $b^\dagger$, and itself. The only
nonvanishing commutant is $[K,b] = -A$, since
$[K,[K,b]] = -[K,A] = 0$, so $U_P b\,U_P^\dagger = b-A$ and
$U_P b^\dagger U_P^\dagger = b^\dagger-A$. Applying $U_P$ to
Eq.~\eqref{eq:H_shifted} in units of $\hbar$, with the qubit term
$\sum_i\omega_{q,i}Z_i/2$ invariant, and using $\Omega A = S$:
\begin{align*}
  U_P(\Omega b^\dagger b)U_P^\dagger &= \Omega b^\dagger b - \Omega A(b+b^\dagger) + \Omega A^2,\\
  U_P[S(b+b^\dagger)]U_P^\dagger    &= S(b+b^\dagger) - 2SA.
\end{align*}
The terms linear in $b+b^\dagger$ cancel exactly. With $SA = S^2/\Omega$ and
$S^2 = (\lambda_1^2+\lambda_2^2)\mathbf{1} + 2\lambda_1\lambda_2 Z_1 Z_2$, the
residual is
$-S^2/\Omega = -(\lambda_1^2+\lambda_2^2)/\Omega\,\mathbf{1} - J\,Z_1 Z_2$,
whose second term is Eq.~\eqref{eq:J_general}.

\section{Magnus Expansion and Exact Propagator}
\label{app:magnus}

The interaction-picture Hamiltonian is
$H_I(t) = \hbar S(b\,e^{-i\Omega t}+b^\dagger e^{i\Omega t})$, and
$U_I(t)=\exp(\mathcal M_1+\mathcal M_2+\cdots)$ with
$\mathcal M_1=-(i/\hbar)\int_0^tH_I\,dt'$ and
$\mathcal M_2=-(1/2\hbar^2)\int_0^tdt_1\int_0^{t_1}dt_2[H_I(t_1),H_I(t_2)]$~\cite{magnus1954exponential,blanes2009magnus}.
The first term is $\mathcal{M}_1(t) = S[\alpha(t)b^\dagger-\alpha^*(t)b]$ with
$\alpha(t) = (1-e^{i\Omega t})/\Omega$. The commutator
$[H_I(t_1),H_I(t_2)] = -2i\hbar^2 S^2\sin[\Omega(t_1-t_2)]$ is a c-number
times $S^2$, which commutes with $H_I(t)$ at all times, so all terms of order
$n\geq 3$ vanish and $\mathcal{M}_2(t) = i\Phi(t)S^2$ with
$\Phi(t) = (\Omega t-\sin\Omega t)/\Omega^2$. Since
$[\mathcal{M}_1,\mathcal{M}_2] = 0$, the propagator factorizes as
Eq.~\eqref{eq:UI_exact}. Because $U_I(t_k)$ depends on $S^2$ alone, the echo
pulses of Sec.~\ref{sec:differential} are applied only at closure times.

At closure, $\alpha(t_k) = 0$ and $\Phi(t_k) = 2\pi k/\Omega^2$, so
$U_I(t_k) = e^{i\Phi(t_k)S^2}$. Expanding $S^2$ gives the global phase
$e^{i\Phi(t_k)(\lambda_1^2+\lambda_2^2)}$ and the entangling operation
$e^{2i\lambda_1\lambda_2\Phi(t_k)Z_1 Z_2} = e^{i\theta_k Z_1 Z_2}$ with
$\theta_k = 4\pi k\lambda_1\lambda_2/\Omega^2 = Jt_k$.

\section{Derivations for the Open-System Dynamics}
\label{app:open}

\subsection{Thermal initial state and reduced qubit dynamics}

The thermal state of Sec.~\ref{subsec:thermal} is
\begin{equation}
  \rho_m^\mathrm{th}(\bar n_0)=\sum_{n=0}^{\infty}p_n|n\rangle\langle n|,
  \qquad
  p_n=\frac{1}{1+\bar n_0}\left(\frac{\bar n_0}{1+\bar n_0}\right)^{\!n}.
  \label{eq:thermal_state}
\end{equation}
The propagator of Eq.~\eqref{eq:UI_exact} acts on the simultaneous eigenstates
$|\bm s\rangle=|s_1,s_2\rangle$ of $Z_1$ and $Z_2$, which diagonalize
$S=\lambda_1Z_1+\lambda_2Z_2$ with eigenvalues
$S_{\bm s}=s_1\lambda_1+s_2\lambda_2$, as
\begin{equation}
  U_I(t)|\bm s\rangle
  = e^{i\Phi(t)S_{\bm s}^2}\,D[\alpha(t)S_{\bm s}]\,|\bm s\rangle .
  \label{eq:prop_on_spin}
\end{equation}
With $\rho_q(0)=\sum_{\bm s,\bm s'}\rho^{(q)}_{\bm s\bm s'}(0)
|\bm s\rangle\langle\bm s'|$ and $\rho_I(t)=U_I(t)\rho(0)U_I^\dagger(t)$,
\begin{align}
  \rho_I(t) ={}& \sum_{\bm s,\bm s'}\rho^{(q)}_{\bm s\bm s'}(0)\,
   e^{i\Phi(t)(S_{\bm s}^2-S_{\bm s'}^2)}\,|\bm s\rangle\langle\bm s'|
   \nonumber\\
   &\otimes\, D(\alpha S_{\bm s})\,\rho_m^\mathrm{th}\,
   D^\dagger(\alpha S_{\bm s'}) .
  \label{eq:full_rho}
\end{align}
Tracing out the oscillator gives
\begin{equation}
  \rho^{(q)}_{\bm s\bm s'}(t)
  = \rho^{(q)}_{\bm s\bm s'}(0)\,
    e^{i\Phi(t)(S_{\bm s}^2-S_{\bm s'}^2)}\,
    \chi_{\bm s\bm s'}(t),
  \label{eq:reduced_element}
\end{equation}
with, by cyclicity of the trace,
\begin{equation}
  \chi_{\bm s\bm s'}(t)
  = \operatorname{Tr}_m\!\left[
      D^\dagger(\alpha S_{\bm s'})D(\alpha S_{\bm s})\,
      \rho_m^\mathrm{th}\right].
  \label{eq:chi_def}
\end{equation}
The Weyl composition relation
$D(\xi_1)D(\xi_2)=\exp[(\xi_1\xi_2^*-\xi_1^*\xi_2)/2]\,D(\xi_1+\xi_2)$ with
$D^\dagger(\xi)=D(-\xi)$ gives
\begin{align}
  D^\dagger(\alpha S_{\bm s'})D(\alpha S_{\bm s})
  ={}& \exp\!\left[\frac{-\alpha S_{\bm s'}\alpha^*S_{\bm s}
        +\alpha^*S_{\bm s'}\alpha S_{\bm s}}{2}\right] \nonumber\\
   &\times D\!\left[\alpha\,\Delta S_{\bm s\bm s'}\right],
  \label{eq:weyl_applied}
\end{align}
where $\Delta S_{\bm s\bm s'}\equiv S_{\bm s}-S_{\bm s'}$. Since
$S_{\bm s}$ and $S_{\bm s'}$ are real the exponent vanishes identically, so
$\chi_{\bm s\bm s'}(t)=\operatorname{Tr}_m[D(\xi)\rho_m^\mathrm{th}]$ with
$\xi=\alpha(t)\Delta S_{\bm s\bm s'}$. Substituting
Eq.~\eqref{eq:thermal_state} and using
$\langle n|D(\xi)|n\rangle=e^{-|\xi|^2/2}L_n(|\xi|^2)$, with $L_n$ the
Laguerre polynomial,
\begin{equation}
  \operatorname{Tr}_m[D(\xi)\rho_m^\mathrm{th}]
  = \frac{e^{-|\xi|^2/2}}{1+\bar n_0}
    \sum_{n=0}^{\infty}
    \left(\frac{\bar n_0}{1+\bar n_0}\right)^{\!n}L_n(|\xi|^2).
  \label{eq:laguerre_sum}
\end{equation}
The generating function
$\sum_{n\ge0}r^nL_n(x)=(1-r)^{-1}\exp[-rx/(1-r)]$, valid for $|r|<1$,
evaluated at $r=\bar n_0/(1+\bar n_0)$, for which $1-r=(1+\bar n_0)^{-1}$ and
$r/(1-r)=\bar n_0$, gives
$\operatorname{Tr}_m[D(\xi)\rho_m^\mathrm{th}]
= e^{-|\xi|^2/2}e^{-\bar n_0|\xi|^2}
= \exp[-\tfrac12(2\bar n_0+1)|\xi|^2]$, which with
$\xi=\alpha(t)\Delta S_{\bm s\bm s'}$ is Eq.~\eqref{eq:thermal_factor}.

\subsection{Conditional oscillator amplitude under damping}

The conditional first moment $\beta_{\bm s}(t)$ of
Sec.~\ref{subsec:damping} follows from Eq.~\eqref{eq:master}. Using
$[b,b^\dagger b]=b$ and $[b,b+b^\dagger]=1$, the Hamiltonian commutator
contributes $-i\Omega\beta_{\bm s}-iS_{\bm s}$ and the two mechanical
dissipators $-(\kappa_m/2)\beta_{\bm s}$, so
\begin{equation}
  \dot\beta_{\bm s}
  = -\left(i\Omega+\frac{\kappa_m}{2}\right)\beta_{\bm s}-iS_{\bm s}.
  \label{eq:beta_ode}
\end{equation}
Subtracting the two instances of Eq.~\eqref{eq:beta_ode} with
$\Delta\beta=\beta_{\bm s}-\beta_{\bm s'}$,
\begin{equation}
  \frac{d}{dt}\Delta\beta
  = -\left(i\Omega+\frac{\kappa_m}{2}\right)\Delta\beta
    -i\,\Delta S_{\bm s\bm s'} .
  \label{eq:dbeta_ode}
\end{equation}
For any initial mechanical state common to both branches,
$\Delta\beta(0)=0$, and the integrating factor
$\exp[(i\Omega+\kappa_m/2)t]$ gives
\begin{equation}
  \Delta\beta(t)
  = -i\,\Delta S_{\bm s\bm s'}\,
    \frac{1-e^{-(i\Omega+\kappa_m/2)t}}{i\Omega+\kappa_m/2},
  \label{eq:dbeta_exact}
\end{equation}
whose squared magnitude is Eq.~\eqref{eq:dbeta_sq}.

\subsection{Information leakage to the bath}

Over an interval $dt$, mechanical damping is a thermal attenuation channel of
transmissivity $\mathcal{T}_{dt}=e^{-\kappa_m dt}=1-\kappa_m dt+O(dt^2)$. In a
Stinespring realization the mode is mixed with an independent thermal bath
mode on a beam splitter, so the amplitudes $\beta_{\bm s}$ and
$\beta_{\bm s'}$ generate bath amplitudes separated by
\begin{equation}
  \delta\zeta = \sqrt{1-{\mathcal{T}_{dt}}}\,\Delta\beta(t)
  = \sqrt{\kappa_m dt}\,\Delta\beta(t)+O(dt^{3/2}).
  \label{eq:bath_separation}
\end{equation}
The bath overlap $\mathcal{V}_{dt}$ is the thermal characteristic function
evaluated above with $\bar n_0\to\bar n_\mathrm{th}$ and
$\xi\to\delta\zeta$:
\begin{equation}
  \mathcal{V}_{dt}
  = \exp\!\left[-\frac{\kappa_m}{2}(2\bar n_\mathrm{th}+1)
      |\Delta\beta(t)|^2\,dt+O(dt^2)\right].
  \label{eq:infinitesimal_visibility}
\end{equation}
Independent bath increments multiply, so
$\prod_j\mathcal{V}_{dt_j}=e^{-\Lambda(t)}$ with $\Lambda(t)$ given by
Eq.~\eqref{eq:Lambda_general}.

On the exact damped path, writing $a=\kappa_m/2$ and $d=\Omega^2+a^2$,
substitution of Eq.~\eqref{eq:dbeta_sq} requires
$\int_0^t d\tau=t$,
$\int_0^t e^{-2a\tau}d\tau=(1-e^{-\kappa_m t})/\kappa_m$, and
$\int_0^t e^{-a\tau}\cos\Omega\tau\,d\tau
=\{a+e^{-at}[-a\cos\Omega t+\Omega\sin\Omega t]\}/d$, giving
\begin{align}
  \Lambda_\mathrm{d}(t) ={}&
   \frac{\kappa_m}{2}\left(2\bar n_\mathrm{th}+1\right)
   \frac{\left(\Delta S_{\bm s\bm s'}\right)^2}{d}
   \Bigg\{t+\frac{1-e^{-\kappa_m t}}{\kappa_m} \nonumber\\
   &-\frac{2}{d}\Big[a+e^{-at}\big(-a\cos\Omega t
     +\Omega\sin\Omega t\big)\Big]\Bigg\}.
  \label{eq:Lambda_exact}
\end{align}
In the weak-damping regime of Sec.~\ref{subsec:damping},
$|\Delta\beta(\tau)|^2$ is replaced by
$|\alpha(\tau)|^2(\Delta S_{\bm s\bm s'})^2$ in
Eq.~\eqref{eq:Lambda_general}. With Eq.~\eqref{eq:alpha_sq},
\begin{equation}
  \int_0^t|\alpha(\tau)|^2d\tau
  = \frac{2}{\Omega^2}\int_0^t\left[1-\cos\Omega\tau\right]d\tau
  = \frac{2}{\Omega^2}\left[t-\frac{\sin\Omega t}{\Omega}\right],
  \label{eq:alpha_sq_integral}
\end{equation}
so the leading-order damping exponent at arbitrary time is
\begin{equation}
  \Lambda(t)
  = \frac{\kappa_m(2\bar n_\mathrm{th}+1)
          \left(\Delta S_{\bm s\bm s'}\right)^2}{\Omega^2}
    \left[t-\frac{\sin\Omega t}{\Omega}\right]
    + O\!\left[\left(\frac{\kappa_m}{\Omega}\right)^{\!2}\right],
  \label{eq:Lambda_weak}
\end{equation}
which at $t_k=2\pi k/\Omega$ reduces to Eq.~\eqref{eq:Lambda_closure}.

\subsection{Qubit dephasing}

Since $Z_i^2=\mathbf 1$, the qubit dissipator is
\begin{equation}
  \frac{\gamma_{\phi,i}}{2}\mathcal{D}[Z_i]\rho
  = \frac{\gamma_{\phi,i}}{2}\left(Z_i\rho Z_i-\rho\right),
  \label{eq:dephasing_explicit}
\end{equation}
which for a computational-basis element $\rho_{\bm s\bm s'}$, with
$Z_i|\bm s\rangle=s_i|\bm s\rangle$, gives
\begin{equation}
  \left.\frac{d}{dt}\rho_{\bm s\bm s'}\right|_{\phi,i}
  = \frac{\gamma_{\phi,i}}{2}\left(s_is_i'-1\right)\rho_{\bm s\bm s'} .
  \label{eq:dephasing_element}
\end{equation}
Integrating and multiplying the factors of the channels with
$s_i\neq s_i'$ gives Eq.~\eqref{eq:dephasing_factor}.

\section{Quantum Fisher Information and Classical Fisher Saturation}
\label{app:qfi}

For the pure output state
$|\psi_\Gamma\rangle = e^{i\theta_k Z_1 Z_2}|{+}{+}\rangle$ the local
generator, defined by $\partial_\Gamma\ket{\psi_\Gamma}=-i\mathcal G\ket{\psi_\Gamma}$,
is $\mathcal{G} = -(\partial_\Gamma\theta_k)Z_1 Z_2$, and
$F_Q=4\,\mathrm{Var}(\mathcal G)$. Since
$\langle{+}{+}|Z_1 Z_2|{+}{+}\rangle = 0$ and $(Z_1 Z_2)^2 = \mathbf{1}$, the
variance is $\mathrm{Var}_{++}(Z_1 Z_2) = 1$, giving
$F_Q = 4(\partial_\Gamma\theta_k)^2$.

The observable $Y_1 Z_2$ anticommutes with $Z_1 Z_2$,
$\{Z_1 Z_2,\,Y_1 Z_2\} = 0$, since $\{Z_1,Y_1\}=0$ and $Z_2^2=\mathbf 1$.
Using $e^{-i\theta \mathcal{A}}\mathcal{B}e^{i\theta \mathcal{A}}
= \mathcal{B}\cos(2\theta)-i\mathcal{A}\mathcal{B}\sin(2\theta)$ for
anticommuting Hermitian $\mathcal{A}$, $\mathcal{B}$ with
$\mathcal{A}^2 = \mathcal{B}^2 = \mathbf{1}$, here $\mathcal{A}=Z_1Z_2$,
$\mathcal{B}=Y_1Z_2$, $\theta=\theta_k$, and
$-i(Z_1 Z_2)(Y_1 Z_2) = -iZ_1 Y_1 = -X_1$ with $X_1$ the Pauli-$x$ operator of
qubit 1:
\begin{equation*}
  \langle Y_1 Z_2\rangle_\Gamma = \langle{+}{+}|[Y_1 Z_2\cos(2\theta_k)-X_1\sin(2\theta_k)]|{+}{+}\rangle = -\sin(2\theta_k).
\end{equation*}
The binary probabilities are $p_\pm = (1\mp\sin(2\theta_k))/2$, and
$\sum_r(\partial_\Gamma p_r)^2/p_r = 4(\partial_\Gamma\theta_k)^2 = F_Q$,
where $r=\pm$ labels the two parity outcomes.

\bibliography{bibliography}

@article{aspelmeyer2014cavity,
  title={Cavity optomechanics},
  author={Aspelmeyer, Markus and Kippenberg, Tobias J and Marquardt, Florian},
  journal={Reviews of Modern Physics},
  volume={86},
  number={4},
  pages={1391--1452},
  year={2014},
  publisher={APS}
}

@article{degen2017quantum,
  title={Quantum sensing},
  author={Degen, Christian L and Reinhard, Friedemann and Cappellaro, Paola},
  journal={Reviews of modern physics},
  volume={89},
  number={3},
  pages={035002},
  year={2017},
  publisher={APS}
}

@article{braunstein1994statistical,
  title={Statistical distance and the geometry of quantum states},
  author={Braunstein, Samuel L and Caves, Carlton M},
  journal={Physical Review Letters},
  volume={72},
  number={22},
  pages={3439},
  year={1994},
  publisher={APS}
}

@article{paris2009quantum,
  title={Quantum estimation for quantum technology},
  author={Paris, Matteo GA},
  journal={International Journal of Quantum Information},
  volume={7},
  number={supp01},
  pages={125--137},
  year={2009},
  publisher={World Scientific}
}

@article{royer2017fast,
  title={Fast and high-fidelity entangling gate through parametrically modulated longitudinal coupling},
  author={Royer, Baptiste and Grimsmo, Arne L and Didier, Nicolas and Blais, Alexandre},
  journal={Quantum},
  volume={1},
  pages={11},
  year={2017},
  publisher={Verein zur F{\"o}rderung des Open Access Publizierens in den Quantenwissenschaften}
}

@article{didier2015fast,
  title={Fast quantum nondemolition readout by parametric modulation of longitudinal qubit-oscillator interaction},
  author={Didier, Nicolas and Bourassa, J{\'e}r{\^o}me and Blais, Alexandre},
  journal={Physical review letters},
  volume={115},
  number={20},
  pages={203601},
  year={2015},
  publisher={APS}
}

@article{bose2017spin,
  title={Spin entanglement witness for quantum gravity},
  author={Bose, Sougato and Mazumdar, Anupam and Morley, Gavin W and Ulbricht, Hendrik and Toro{\v{s}}, Marko and Paternostro, Mauro and Geraci, Andrew A and Barker, Peter F and Kim, MS and Milburn, Gerard},
  journal={Physical review letters},
  volume={119},
  number={24},
  pages={240401},
  year={2017},
  publisher={APS}
}

@article{marletto2017gravitationally,
  title={Gravitationally induced entanglement between two massive particles is sufficient evidence of quantum effects in gravity},
  author={Marletto, Chiara and Vedral, Vlatko},
  journal={Physical review letters},
  volume={119},
  number={24},
  pages={240402},
  year={2017},
  publisher={APS}
}

@article{torovs2024loss,
  title={Loss of coherence and coherence protection from a graviton bath},
  author={Toro{\v{s}}, Marko and Mazumdar, Anupam and Bose, Sougato},
  journal={Physical Review D},
  volume={109},
  number={8},
  pages={084050},
  year={2024},
  publisher={APS}
}

@article{miki2024quantum,
  title={Quantum signature of gravity in optomechanical systems with conditional measurement},
  author={Miki, Daisuke and Matsumura, Akira and Yamamoto, Kazuhiro},
  journal={Physical Review D},
  volume={109},
  number={6},
  pages={064090},
  year={2024},
  publisher={APS}
}

@article{gonzalezballestero2021levitodynamics,
  author  = {Gonzalez-Ballestero, C. and Aspelmeyer, M. and Novotny, L. and Quidant, R. and Romero-Isart, O.},
  title   = {Levitodynamics: {L}evitation and control of microscopic objects in vacuum},
  journal = {Science},
  volume  = {374},
  pages   = {eabg3027},
  year    = {2021},
  doi     = {10.1126/science.abg3027}
}

@article{westphal2021measurement,
  author  = {Westphal, Tobias and Hepach, Hans and Pfaff, Jeremias and Aspelmeyer, Markus},
  title   = {Measurement of gravitational coupling between millimetre-sized masses},
  journal = {Nature},
  volume  = {591},
  pages   = {225--228},
  year    = {2021},
  doi     = {10.1038/s41586-021-03250-7}
}

@article{oconnell2010quantum,
  author  = {O'Connell, A. D. and Hofheinz, M. and Ansmann, M. and Bialczak, Radoslaw C. and Lenander, M. and Lucero, Erik and Neeley, M. and Sank, D. and Wang, H. and Weides, M. and Wenner, J. and Martinis, John M. and Cleland, A. N.},
  title   = {Quantum ground state and single-phonon control of a mechanical resonator},
  journal = {Nature},
  volume  = {464},
  pages   = {697--703},
  year    = {2010},
  doi     = {10.1038/nature08967}
}

@article{kasevich1991atomic,
  author  = {Kasevich, Mark and Chu, Steven},
  title   = {Atomic interferometry using stimulated {R}aman transitions},
  journal = {Phys. Rev. Lett.},
  volume  = {67},
  pages   = {181--184},
  year    = {1991},
  doi     = {10.1103/PhysRevLett.67.181}
}

@article{qvarfort2018gravimetry,
  author  = {Qvarfort, Sofia and Serafini, Alessio and Barker, P. F. and Bose, Sougato},
  title   = {Gravimetry through non-linear optomechanics},
  journal = {Nat. Commun.},
  volume  = {9},
  pages   = {3690},
  year    = {2018},
  doi     = {10.1038/s41467-018-06037-z}
}

@article{armata2017quantum,
  author  = {Armata, F. and Latmiral, L. and Plato, A. D. K. and Kim, M. S.},
  title   = {Quantum limits to gravity estimation with optomechanics},
  journal = {Phys. Rev. A},
  volume  = {96},
  pages   = {043824},
  year    = {2017},
  doi     = {10.1103/PhysRevA.96.043824}
}

@article{giovannetti2011advances,
  author  = {Giovannetti, Vittorio and Lloyd, Seth and Maccone, Lorenzo},
  title   = {Advances in quantum metrology},
  journal = {Nat. Photonics},
  volume  = {5},
  pages   = {222--229},
  year    = {2011},
  doi     = {10.1038/nphoton.2011.35}
}

@article{matsumura2020gravity,
  author  = {Matsumura, Akira and Yamamoto, Kazuhiro},
  title   = {Gravity-induced entanglement in optomechanical systems},
  journal = {Phys. Rev. D},
  volume  = {102},
  pages   = {106021},
  year    = {2020},
  doi     = {10.1103/PhysRevD.102.106021}
}

@article{krisnanda2020observable,
  author  = {Krisnanda, Tanjung and Tham, Guo Yao and Paternostro, Mauro and Paterek, Tomasz},
  title   = {Observable quantum entanglement due to gravity},
  journal = {npj Quantum Inf.},
  volume  = {6},
  pages   = {12},
  year    = {2020},
  doi     = {10.1038/s41534-020-0243-y}
}

@article{ali2024quantum,
  author        = {Ali, Asad and Al-Kuwari, Saif and Ghominejad, Mehrdad and Rahim, M. T. and Wang, Dong and Haddadi, Saeed},
  title         = {Quantum characteristics near event horizons},
  journal       = {Phys. Rev. D},
  volume        = {110},
  pages         = {064001},
  year          = {2024},
  doi           = {10.1103/PhysRevD.110.064001},
  eprint        = {2401.12028},
  archivePrefix = {arXiv}
}

@article{will2014confrontation,
  author  = {Will, Clifford M.},
  title   = {The confrontation between general relativity and experiment},
  journal = {Living Rev. Relativ.},
  volume  = {17},
  pages   = {4},
  year    = {2014},
  doi     = {10.12942/lrr-2014-4}
}

@article{hahn1950spin,
  author  = {Hahn, E. L.},
  title   = {Spin echoes},
  journal = {Phys. Rev.},
  volume  = {80},
  pages   = {580--594},
  year    = {1950},
  doi     = {10.1103/PhysRev.80.580}
}

@article{sorensen2000entanglement,
  author  = {S{\o}rensen, Anders and M{\o}lmer, Klaus},
  title   = {Entanglement and quantum computation with ions in thermal motion},
  journal = {Phys. Rev. A},
  volume  = {62},
  pages   = {022311},
  year    = {2000},
  doi     = {10.1103/PhysRevA.62.022311}
}

@article{leibfried2003experimental,
  author  = {Leibfried, D. and DeMarco, B. and Meyer, V. and Lucas, D. and Barrett, M. and Britton, J. and Itano, W. M. and Jelenkovi{\'c}, B. and Langer, C. and Rosenband, T. and Wineland, D. J.},
  title   = {Experimental demonstration of a robust, high-fidelity geometric two ion-qubit phase gate},
  journal = {Nature},
  volume  = {422},
  pages   = {412--415},
  year    = {2003},
  doi     = {10.1038/nature01492}
}

@article{magnus1954exponential,
  author  = {Magnus, Wilhelm},
  title   = {On the exponential solution of differential equations for a linear operator},
  journal = {Commun. Pure Appl. Math.},
  volume  = {7},
  pages   = {649--673},
  year    = {1954},
  doi     = {10.1002/cpa.3160070404}
}

@article{blanes2009magnus,
  author  = {Blanes, S. and Casas, F. and Oteo, J. A. and Ros, J.},
  title   = {The {M}agnus expansion and some of its applications},
  journal = {Phys. Rep.},
  volume  = {470},
  pages   = {151--238},
  year    = {2009},
  doi     = {10.1016/j.physrep.2008.11.001}
}

@book{misner1973gravitation,
  author    = {Misner, Charles W. and Thorne, Kip S. and Wheeler, John Archibald},
  title     = {Gravitation},
  publisher = {W. H. Freeman},
  address   = {San Francisco},
  year      = {1973}
}

@article{wootters1998entanglement,
  title={Entanglement of formation of an arbitrary state of two qubits},
  author={Wootters, William K},
  journal={Physical Review Letters},
  volume={80},
  number={10},
  pages={2245},
  year={1998},
  publisher={APS}
}

@article{ali2024trade,
  title={Trade-off relations of quantum resource theory in Heisenberg models},
  author={Ali, Asad and Al-Kuwari, Saif and Haddadi, Saeed},
  journal={arXiv preprint arXiv:2401.01063},
  year={2024}
}

@article{ali2024ergotropy,
  title={Ergotropy and capacity optimization in Heisenberg spin-chain quantum batteries},
  author={Ali, Asad and Al-Kuwari, Saif and Hussain, MI and Byrnes, Tim and Rahim, MT and Quach, James Q and Ghominejad, Mehrdad and Haddadi, Saeed},
  journal={Physical Review A},
  volume={110},
  number={5},
  pages={052404},
  year={2024},
  publisher={APS}
}

@article{bengyat2024,
  title = {Gravity-mediated entanglement between oscillators as quantum superposition of geometries},
  author = {Bengyat, Ofek and Di Biagio, Andrea and Aspelmeyer, Markus and Christodoulou, Marios},
  journal = {Phys. Rev. D},
  volume = {110},
  issue = {5},
  pages = {056046},
  numpages = {8},
  year = {2024},
  month = {Sep},
  publisher = {American Physical Society},
  doi = {10.1103/PhysRevD.110.056046},
  url = {https://link.aps.org/doi/10.1103/PhysRevD.110.056046}
}

@article{tang2025,
  title = {Optimal form factors for experimental proposals on gravity-induced entanglement},
  author = {Tang, Ziqian and Xue, Hanyu and Han, Zizhao and Kan, Zikuan and Li, Zeji and Liu, Yulong},
  journal = {Phys. Rev. D},
  volume = {112},
  issue = {4},
  pages = {042004},
  numpages = {13},
  year = {2025},
  month = {Aug},
  publisher = {American Physical Society},
  doi = {10.1103/jznw-q3q8},
  url = {https://link.aps.org/doi/10.1103/jznw-q3q8}
}

@article{hussain2014geometric,
  title={Geometric phase gate for entangling two Bose-Einstein condensates},
  author={Hussain, Mahmood Irtiza and Ilo-Okeke, Ebubechukwu O and Byrnes, Tim},
  journal={Physical Review A},
  volume={89},
  number={5},
  pages={053607},
  year={2014},
  publisher={APS},
   doi = {10.1103/PhysRevA.89.053607},
  url = {https://journals.aps.org/pra/abstract/10.1103/PhysRevA.89.053607}
}

@article{hussain2015geometric,
  title={Geometric phase gate based on the ac Stark shift},
  author={Hussain, Mahmood Irtiza and Ilo-Okeke, Ebubechukwu O and Byrnes, Tim},
  journal={Quantum Information Processing},
  volume={14},
  number={3},
  pages={943--950},
  year={2015},
  publisher={Springer},
  doi={10.1007/s11128-014-0907-7}
}

\end{document}